\documentclass[fleqn,usenatbib]{mnras}
\usepackage{newtxtext,newtxmath}
\usepackage{comment}
\usepackage{multirow}
\usepackage{mathtools}
\usepackage[T1]{fontenc}
\usepackage{ae,aecompl}

\usepackage{graphicx}	
\usepackage{amsmath}	
\usepackage[ruled,vlined]{algorithm2e}
\usepackage{algpseudocode}
\usepackage{enumitem} 
\usepackage{xfrac} 

\newcommand{\mdot}{$\mathrm{M}_\odot$}
\newcommand{\zdot}{$\mathrm{Z}_\odot$}
\newcommand{\ldot}{$\mathrm{L}_\odot$}
\newcommand{\sphgal}{\textsc{sphgal}}
\newcommand{\gadget}{\textsc{gadget-3}}
\newcommand{\healpix}{\textsc{healpix}}
\newcommand{\treecol}{\textsc{treecol}}
\newcommand{\bifrost}{\textsc{bifrost}}
\newcommand{\ketju}{\textsc{ketju}}
\newcommand{\mstar}{\textsc{mstar}}
\newcommand{\griffin}{\textsc{griffin}}
\newcommand{\boost}{\textsc{BoOST}}
\newcommand{\skirt}{\textsc{skirt}}

\title[The structure of a young massive star cluster]{Mergers all the way down: stellar collisions and kinematics of a dense hierarchically forming massive star cluster in a dwarf starburst}
\author[Lahén et al.]{Natalia Lahén$^{1}$\thanks{E-mail: nlahen@mpa-garching.mpg.de}, Thorsten Naab$^{1}$, Antti Rantala$^{1}$ and Christian Partmann$^{1,2}$\\
$^{1}$Max-Planck-Institut f\"ur Astrophysik, Karl-Schwarzschild-Str. 1, 
D-85748, Garching, Germany\\
$^{2}$Center for Computational Astrophysics, Flatiron Institute, 162 5th Avenue, New York, NY 10010, US\\
}
\date{Accepted XXX. Received YYY; in original form ZZZ}
\pubyear{2025}

\begin{document}
\label{firstpage}
\pagerange{\pageref{firstpage}--\pageref{lastpage}}
\maketitle

\begin{abstract}

Recent observations indicate that the progenitors of globular clusters (GCs) at high redshifts had high average stellar surface densities above \mbox{$10^5$ M$_\odot$pc$^{-2}$}. The internal structure and kinematics of the clusters, however, remain out of reach. Numerical simulations are necessary to decipher the origin of spatio-kinematic features in present-day GCs. Here we study star cluster formation in a star-by-star hydrodynamical simulation of a low-metallicity starburst in a merger of two gas-rich dwarf galaxies. The simulation accounts for the multiphase interstellar medium, stellar radiation, winds and supernovae, and the accurate small-scale gravitational dynamics near massive stars. We also include prescriptions for stellar collisions and tidal disruption events by black holes. Gravitationally bound star clusters up to \mbox{$\sim2\times10^5$ M$_\odot$} form dense with initial half-mass radii of $\sim0.1$–\mbox{1 pc}. The most massive cluster approaches the observed high-redshift surface densities throughout its hierarchical and dissipative assembly. The cluster also hosts a collisionally growing very massive star of \mbox{$\sim1000$ M$_\odot$} that will eventually collapse, forming an intermediate mass black hole. The assembly leaves an imprint in the spatio-kinematic structure of the cluster. The youngest stars are more centrally concentrated, they show significant bulk rotation and have radially biased velocity components at outer radii. The older population is more round in shape, rotates slowly, its velocity distribution is isotropic and exhibits higher dispersion. If chemically enriched star formation proceeds mainly in the later stages of cluster assembly, these results provide a possible explanation for some of the multiple population features observed in dynamically young GCs.

\end{abstract}

\begin{keywords}
galaxies: dwarf  -- galaxies: star clusters: general -- galaxies: star formation -- gravitation -- methods: numerical -- stars: massive
\end{keywords}


\section{Introduction}

Young massive star clusters are among the densest gravitationally bound stellar structures \citep{2005ApJS..161..304M, 2010ARA&A..48..431P, 2011MNRAS.414.3699M}. The most extreme examples are super star clusters in local starburst galaxies such as the M82 \citep{2007ApJ...663..844M, 2024ApJ...973L..55L}, and young massive star clusters observed through gravitational lensing in the high-redshift Universe, such as in the Sunrise Arc and the Cosmic Gems \citep{2023ApJ...945...53V, 2024Natur.632..513A}. In both cases, the highest resolved effective surface densities ($\Sigma_\mathrm{eff}$) within the effective radius ($R_\mathrm{eff}$; projected half-mass or half-light radius) exceed \mbox{$10^5$ \mdot{} pc$^{-2}$}. The smallest $R_\mathrm{eff}$ of these massive ($M_\mathrm{cluster}$\mbox{$>10^5$ \mdot{}}) clusters are less than \mbox{1 pc}. Runaway stellar collisions and black hole growth \citep{Gold1965,Spitzer1966,2004Natur.428..724P} as well as chemical self-enrichment through gas retention \citep{ 2016A&A...587A..53K, 2018A&A...615A.119V} have been suggested to take place at such high densities. However, the details of the internal structure of the densest clusters still remain unresolved.

In nearby systems, where the internal structure of clusters can be resolved down to individual stars, clusters that are both young and massive are rare. The R136 cluster in the Large Magellanic Cloud (LMC) is one of the best examples with an age of \mbox{$\sim 1.5$ Myr} \citep{2016MNRAS.458..624C}. The total mass of the cluster has been estimated as $>2\times 10^4$ \mdot{} within the central few pc or up to \mbox{$\sim 10^5$ \mdot{}} in total \citep{2009ApJ...707.1347A}. The $\Sigma_\mathrm{eff}$ of R136 is, however, only a few hundred \mbox{\mdot{} pc$^{-2}$}. Other notable young massive clusters in the Local Group are NGC 3603 \citep{2008ApJ...675.1319H}, Westerlund 1 \citep{2013AJ....145...46L}, Westerlund 2 \citep{2007A&A...466..137A} and the Arches cluster \citep{2009A&A...501..563E} which have mostly higher metallicities than R136 ($Z\sim$ \mbox{$0.5$ \zdot}). None of these clusters reach $\Sigma_\mathrm{eff}$ as high as the starburst or high-redshift massive clusters.

Present-day globular clusters (GCs) in the Local Group can also be used to infer some details of the initial properties of their massive progenitors. This is however not straightforward, as GCs undergo mass segregation, mass loss and angular momentum transport (in case they rotate), through stellar evolution, relaxation and tidal effects during the several Gyrs of evolution. While the two-body relaxation time can be short in the central region of an initially dense cluster, the more sparse outer parts can still retain some of the spatio-kinematic information that was imprinted during the assembly of the cluster \citep{2013MNRAS.429.1913V, 2015MNRAS.450.1164H, 2017MNRAS.469..683T}. \citet{2018MNRAS.481.2125B}, \citet{2023A&A...671A.106M} and \citet{2025A&A...694A.184L} for instance find that a significant number of GCs rotate. The rotation speed over velocity dispersion ($V/\sigma$) is observed to correlate with the mass and the relaxation time of the clusters. If the less dynamically evolved systems rotate more, then the angular momentum may indeed be primordial rather than the result of, for instance, later tidal interactions. Hydrodynamical simulations of cluster formation (e.g. \citealt{2017MNRAS.467.3255M}) support these findings, showing that increasingly massive clusters inherit a larger specific angular momentum from the progenitor gas cloud \citep{2020ApJ...904...71L}. Clusters forming in filamentary complexes can form with anisotropic velocity distributions, for instance showing radial bias in the outer parts \citep{2025ApJ...984...75K}.

The spatial and kinematic differences of the stellar populations in present-day GCs provide additional information of the initial stellar distribution. Multiple populations (MPs) are identified through stellar light-element variations that are thought to originate in the early stages of the GC evolution and may thus carry information of the GC formation process (\citealt{2001A&A...369...87G, 2009A&A...505..139C, 2015AJ....149...91P, 2017A&A...601A.112P, 2019MNRAS.487.3815M, 2025MNRAS.537.2342C}; see e.g. \citealt{2015MNRAS.454.4197R, 2016EAS....80..177C, 2018ARA&A..56...83B, 2019A&ARv..27....8G, 2022Univ....8..359M} for reviews). \citet{2019ApJ...884L..24D} found that the light-element enhanced second population (P2) in dynamically young GCs tends to be more centrally concentrated than the primordial, field-like population (P1). \citet{2023MNRAS.520.1456L} confirmed the possibility of centrally concentrated P2 but also found a number of clusters for which P1 is instead centrally concentrated. The majority of their clusters, however, showed fully mixed populations. In addition, \citet{2024A&A...691A..94D} found that P2 has predominantly a higher $V/\sigma$ compared to the P1 in the majority of their analysed clusters. This scenario is supported by hydrodynamical simulations of the formation of the P2 through gas accretion and/or winds of evolved intermediate mass stars in a pre-existing cluster of P1 stars \citep{2010ApJ...724L..99B, 2021MNRAS.500.4578M, 2022MNRAS.517.1171L}. \citet{2023A&A...671A.106M} and \citet{2025A&A...694A.184L}, on the other hand, only found one or two such clusters in their samples of 25 and 30 GCs and found no differences in the kinematics of P1 and P2 in the majority of their GCs. Finally, the P1 and P2 sometimes exhibit different velocity anisotropies \citep{2013ApJ...771L..15R, 2015ApJ...810L..13B, 2025MNRAS.537.2342C}. N-body simulations indicate that primordial spatial or kinematic differences may be lost during the evolution of the GC in a galactic tidal field except for the most dynamically young clusters \citep{2008A&A...492..101D}, and the rate of this relaxation depends on the initial velocity and anisotropy properties \citep{2024A&A...689A.313P}.

A self-consistent numerical approach to the initial properties of the populations of stars in young massive clusters forming in a realistic galactic environment is therefore needed. An increasing variety of numerical simulations is now able to probe a wide range of interstellar environments that give rise to the formation of massive star clusters \citep{2015MNRAS.446.2038R, 2020MNRAS.493.4315M, 2020ApJ...891....2L, 2021PASJ...73.1036H, 2022MNRAS.509..954D, 2023MNRAS.519.1366G, 2023MNRAS.522.1800S, 2025ApJ...981L..28M, 2025A&A...698A.207C, 2024A&A...681A..28A, 2024ApJ...971..103G, 2025ApJ...978...15R, 2025ApJ...990..135W, 2025PASJ..tmp...12M}. Both observations and simulations agree that galactic systems exhibiting high interstellar gas pressures and high star formation densities harbour the most massive star clusters \citep{2017ApJ...839...78J, 2020MNRAS.499.3267A, 2020ApJ...891....2L}. However, because dense star clusters are inherently collisional systems, models of their formation and internal evolution should account for stellar two and few body interactions (e.g. \citealt{Aarseth2003, 2003gmbp.book.....H}). A number of recent hydrodynamical studies have explored star cluster formation and/or evolution at such high fidelity: \citet{2021PASJ...73.1074F}, \citet{2022MNRAS.512..216G}, \citet{2024MNRAS.527.6732F}, \citet{2024Sci...384.1488F}, \citet{2024A&A...690A..94P} and \citet{2024ApJ...977..203C} modelled star cluster formation in individual molecular clouds at very high spatial resolution and including models for stellar evolution. \citet{2024Sci...384.1488F}, \citet{2024A&A...690A..94P} and \citet{2024ApJ...977..203C} showed that massive star clusters can exhibit very high densities and star formation efficiencies, in agreement or even exceeding the high-$z$ surface densities, when the clusters are still embedded. Cloud-scale simulations are, however, limited to a low number of clusters. They do not consider the role of the galactic environment and thus the impact of the galactic tidal field, gaseous inflows and outflows and the complex time and space varying distribution of feedback from other nearby star forming regions. Toward this end, \citet{2024ApJ...974..193J} introduced a method for modelling the dynamics of stars in a galaxy-scale hydrodynamical simulation but did not include a realistic stellar mass function or star formation. 

In \citet{2025MNRAS.538.2129L} we presented the first star-by-star hydrodynamical simulations of galaxy-scale star cluster formation that accounted for collisional stellar dynamics in gravitational interactions that occur near massive stars. We used the \ketju{} integration module  \citep{2017ApJ...840...53R, 2023MNRAS.524.4062M} incorporated in the \sphgal{} galaxy evolution code \citep{2005MNRAS.364.1105S, 2014MNRAS.443.1173H, 2016MNRAS.458.3528H, 2017MNRAS.471.2151H, 2023MNRAS.522.3092L, 2025MNRAS.537..956P}. These simulations followed the formation of relatively low mass (up to $\sim 1000$ \mdot) and low density ($\Sigma_\mathrm{eff}\lesssim $\mbox{$10^4$ \mdot{} pc$^{-2}$}) clusters in a quiescent low-metallicity dwarf galaxy environment. The clusters formed with compact, sub-parsec half-mass radii and rapidly expanded onto the observed mass-size relation over the first 10 Myr. In the present study we update and apply the numerical methods presented in \citet{2025MNRAS.538.2129L} in an extreme starburst environment of a gas-rich dwarf galaxy merger. We investigate the initial structure and kinematics of massive star clusters star-by-star in a self-consistent low-metallicity galactic environment.

The paper is structured as follows. Section \ref{section:methods} gives a brief overview of the hydrodynamical simulation code, newly implemented methods, initial conditions and post-processing techniques. Section \ref{section:results} describes the results of the starburst simulation, including a description of the mass-size and size-surface density distribution of star clusters. We also detail the stellar collisions that occur in the cores of the dense clusters, and the growth of the most massive star in the simulation. We describe the internal density and kinematic properties of the most massive cluster
and its stellar populations in Section \ref{section:MMC}. Section \ref{section:conclusions} provides conclusions and a discussion of the main results.

\section{Simulations}\label{section:methods}

The starburst simulation presented here is a part of the Galaxy Realizations Including Feedback From INdividual massive stars\footnote{\url{https://wwwmpa.mpa-garching.mpg.de/~naab/griffin-project}} (\textsc{griffin}) project. Details of the numerical methods have previously been described in \citet{2014MNRAS.443.1173H, 2016MNRAS.458.3528H, 2017MNRAS.471.2151H, 2019MNRAS.483.3363H, 2023MNRAS.522.3092L, 2025MNRAS.538.2129L, 2025MNRAS.537..956P} and we provide a brief summary of the main features of our code here.  

\subsection{Hydrodynamical simulations}

We use the smoothed particle hydrodynamics (SPH) code \sphgal{} \citep{2014MNRAS.443.1173H, 2016MNRAS.458.3528H, 2017MNRAS.471.2151H} that is an updated version of the \gadget{} code \citep{2005MNRAS.364.1105S} with improvements to the numerical and algorithmic accuracy of the hydrodynamical methods. The code solves the non-equilibrium cooling and heating processes at low temperatures ($<3\times 10^4$ K) using a chemical network (H$_2$, H$^+$, H, CO, C$^+$, O and free electrons) following the methods of \citet{1997ApJ...482..796N}, \citet{2007ApJS..169..239G} and \citet{2012MNRAS.421..116G}. At higher temperatures tabulated metallicity dependent cooling rates from \citet{2009MNRAS.393...99W} are used. 

\subsubsection{Star formation}

Star formation is triggered when the gas conditions exceed a threshold given by the local Jeans mass $M_\mathrm{J}$ set as $M_\mathrm{J}<0.5 M_\mathrm{SPH}$ where $m_\mathrm{SPH}$ is the SPH kernel mass ($\sim 400$ \mdot) and 
\begin{equation}
   M_\mathrm{J} = \frac{\pi^{5/2}c_\mathrm{s}^3}{6G^{3/2}\rho^{1/2}},
\end{equation}
$c_\mathrm{s}$ being the sound speed, $\rho$ the gas density and $G$ is the gravitational constant. Any particle exceeding the threshold is instantaneously turned into a star particle with a mass and metallicity of the gas particle. The new star particles then wait for the local dynamical time scale $t_\mathrm{dyn}=(4\pi G \rho)^{-1/2}$ after which they are sampled into individual stars along the \citet{2001MNRAS.322..231K} IMF between \mbox{0.08 \mdot{}} and \mbox{500 \mdot}. 

Conservation of mass is enforced in the sampling process within the Jeans length 
\begin{equation}
   R_\mathrm{J} = \left( \frac{3}{4\pi} \frac{M_\mathrm{J}}{\rho} \right) ^{1/3},
\end{equation}
measured when the particle was turned into a star particle. The mass of the star particle (fiducial gas mass resolution \mbox{$\sim 4$ \mdot{}}) is sampled into several stars and the last sampled star exceeding the particle mass is taken from other such particles waiting to be sampled within $R_\mathrm{J}$. This ensures that very massive stars (VMSs; $\gtrsim100$ \mdot) can only form in relatively dense star-forming regions as demonstrated in \citealt{2023MNRAS.522.3092L} and \citet{2024MNRAS.530..645L}. The most massive star formed this way reaches \mbox{$\sim330$ \mdot}, and more massive stars therefore form only through stellar collisions (Section \ref{section:collisions}). New stars are placed around the parent particle in a Gaussian position and velocity distribution with standard deviation equal to \mbox{0.1 pc} and \mbox{0.1 km s$^{-1}$} respectively in each cartesian direction.  Each star is initialised as a single star, and we do not consider primordial binaries in this study. To prevent the creation of tightly bound initial binaries, we resample positions and velocities of stars that are bound to another new star within a semimajor axis of less than \mbox{$5\times 10^{-3}$ pc}. We also check at the end of the routine that no star is placed closer than \mbox{$10^{-3}$ pc} to another new star, and shift the positions of such stars by an additional Gaussian \mbox{0.01 pc}. Random initialisation of such binaries is however negligibly rare.

\subsubsection{Stellar feedback}

We consider mass and energy release through radiation, stellar winds and supernovae. Events that we consider to occur instantaneously at the end of the stellar lifetime are core-collapse supernovae (zero age main sequence mass $8$--$40$ \mdot{}, \citealt{2004ApJ...608..405C}), pair-instability supernovae ($107.2$--\mbox{$203.4$ \mdot{}}; according to the exploding helium core masses from \citealt{2002ApJ...567..532H}) and asymptotic giant branch winds ($0.5$--\mbox{$8$ \mdot{}}, \citealt{2010MNRAS.403.1413K}). The stellar lifetimes are estimated according to the Geneva stellar models \citep{2013A&A...558A.103G} for low-mass stars ($<9$ \mdot) and using the Bonn Optimized Stellar Tracks (\boost; \citealt{2022A&A...658A.125S}) for massive stars ($9$--\mbox{$500$ \mdot{}}). Winds of massive stars (\mbox{$>9$ \mdot{}}) are treated with a continuous momentum-conserving approach. The wind mass loss rates and velocities depend on the initial stellar mass and age, interpolated using the \boost models at metallicity $Z=$\mbox{$0.01$ \zdot}. We further describe in Section \ref{section:collisions} our treatment of stellar mergers and stellar masses that exceed the upper mass limit of the tables. Chemical enrichment is followed by tracking the mass of 13 chemical elements per particle: H, He, N, C, O, Si, Al, Na, Mg, Fe, S, Ca and Ne. 

Each star contributes to the interstellar radiation field in the far-ultraviolet wavelength range of \mbox{$6$--$13.6$ eV}. The fixed zero age main sequence luminosities for stars below $9$ \mdot{} are integrated from the \textsc{BaSeL} spectral library at $Z\sim 0.01$ \zdot{} \citep{1997A&AS..125..229L, 1998A&AS..130...65L, 2002A&A...381..524W} using the \textsc{geneva} stellar models at $Z=0.0004\sim 0.02$ \zdot \citep{2019A&A...627A..24G}\footnote{For stars of \mbox{$<1.7$ \mdot} that are not tabulated in the $Z=0.0004$\mbox{$\sim 0.02$ \zdot{}} Geneva tables we scale the fluxes from the $Z=0.002$\mbox{$\sim 0.1$ \zdot{}} tables of \citet{2013A&A...558A.103G} between \mbox{$0.8$--$1.7$ \mdot{}} by a factor of 2. For stars below \mbox{$0.8$ \mdot{}} we extrapolate as $L\propto M^{3.5}$.}. Massive stars ($>9$ \mdot{}) are given initial mass and age dependent rates of far-ultraviolet and hydrogen ionizing photons ($>13.6$ eV) from the \boost{} tables.

The strength of the interstellar radiation field is computed at the location of each gas particle using the \treecol-algorithm \citep{2012MNRAS.420..745C}. The incoming flux is attenuated by gas and dust column density assuming optically thin conditions in 12 solid angles divided using the \healpix{} algorithm \citep{2011ascl.soft07018G}. For injection of supernovae and winds we likewise use \healpix{} to distribute the $1/12$th fraction of mass and energy equally into $8\pm2$ closest gas particles in each solid angle. Photoionization is implemented as an approximation wherein the photoionized gas within the Str\"omgren sphere is set fully ionized at $10^4$ K and the ionization equilibrium is solved iteratively when HII regions overlap. More details for the feedback implementation can be found in \citet{2016MNRAS.458.3528H, 2017MNRAS.471.2151H} and \citet{2023MNRAS.522.3092L}.

\subsubsection{Collisional stellar gravitational dynamics}\label{section:ketju}

In \citet{2025MNRAS.537..956P} and \citet{2025MNRAS.538.2129L} we described how to combine \sphgal{} with the \ketju{} dynamics module \citep{2017ApJ...840...53R, 2023MNRAS.524.4062M}. All stellar gravitational interactions within a radius $r_\mathrm{KETJU}$ of any star or a remnant above a chosen initial mass limit $m_i$ are treated with the algorithmically regularised integrator \mstar{} \citep{2020MNRAS.492.4131R}. The regularised integration in the \ketju{} regions enables accurate calculation of the stellar trajectories without the use of gravitational softening. The accuracy parameters in \ketju{} are the same as in \citet{2025MNRAS.538.2129L}: Gragg-Bulirsch-Stoer tolerance is set to $\eta_\mathrm{GBS}=10^{-7}$ and the end-time iteration tolerance is $\eta_\mathrm{t}=10^{-3}$.

The computational cost of \ketju{} scales with both the number of regions (linear scaling) and the number of particles in the regions (near quadratic scaling), therefore both $r_\mathrm{KETJU}$ and $m_i$ have to be selected optimally for the specific problem. The size of the \ketju-integration region has to be set as at least $2.8$ times the gravitational softening length ($\epsilon_*$) used in the standard tree force calculation of \gadget. Here we use \mbox{$\epsilon_*=0.01$ pc} and therefore \mbox{$r_\mathrm{KETJU}=0.03$ pc}. This value of $\epsilon_*$ is selected to give a manageable number of up to a few hundred stars in the \ketju-regions within massive clusters that can reach more than $10^6$ stars per pc$^3$ in our simulations. In practice, overlapping \ketju-regions are combined into one, therefore the cores of the young massive clusters will have somewhat larger regularised regions as long as massive stars are not expelled. In the most massive cluster, up to a few $10^4$ stars are within the central \ketju-region that extends to a radius of $0.1$ pc within the centre of mass. 

As the present study concentrates on the most massive clusters that are dense and host a large number of massive stars, we set $m_i=8$ \mdot. This guarantees that the dynamics of massive stars, their remnants and their neighbours in the cores of most of the young clusters more massive than a few hundred \mdot{} will be regularised (see the results of IMF sampling in \citealt{2023MNRAS.522.3092L}). This results in a significantly improved dynamical evolution of the stellar component in the clusters as demonstrated in \citet{2025MNRAS.538.2129L} via comparison to direct N-body simulations performed with the \bifrost{} code \citep{2023MNRAS.522.5180R}.

\subsubsection{Stellar collisions}\label{section:collisions}

At high stellar densities, stellar collisions can lead to the growth of massive objects in the cores of the clusters (see e.g. N-body and cloud-scale simulations of \citealt{2002ApJ...576..899P, 2004Natur.428..724P, 2021MNRAS.501.5257R, 2024Sci...384.1488F, 2024MNRAS.531.3770R}). Following stellar evolution combined with interactions and mergers in a sub-resolution model would require understanding how such mergers proceed (e.g. mixing and mass loss, \citealt{2002ApJ...568..939L, 2013MNRAS.434.3497G}) and how the remnants of the interactions evolve \citep{2009A&A...497..255G, 2013ApJ...764..166D, 2024ApJ...963L..42M}. In the absence of comprehensive models for stellar evolution under repeated collisions we thus implement a mass conserving stellar collision approximation following ideas that are often applied in rapid binary population synthesis (e.g. \citealt{2000MNRAS.315..543H}; see \citealt{2020MNRAS.497.4549A} and \citealt{2023MNRAS.524..426I} for modern approaches). 

We assume that a collision occurs when the radii of two stars overlap. N-body simulations indicate that there is usually only one massive object growing in the cluster core through collisions and that the majority of the mass gain comes in the form of other massive stars \citep{2024MNRAS.531.3770R}. We use the stellar radii provided in the \boost{} models ($m_i>9$ \mdot) to track the collisions, therefore collisions can only occur when one of the progenitors is a massive star. When a collision with another star occurs, the current mass of the less massive star is added to the more massive one and the age over lifetime ratio $f_\mathrm{age}=\mathrm{age}/\mathrm{lifetime}$ for the more massive star is recorded. We then go through \boost{} stellar models with increasing initial mass to find an epoch that first matches the total (current) mass of the star and the fractional age $f_\mathrm{age}$ on the specific track. As a conservative estimate of the total lifetime of the star, we only update the current age of the star to match the respective fractional age on the new higher mass stellar track. We thus do not implement any explicit rejuvenation methods unlike is often done in N-body models that consider mass transfer or collisions \citep{1997MNRAS.291..732T, 1999A&A...348..117P}. We will further refine the assignment of the new stellar age in the future, to account for the evolutionary stage (e.g. to retain the core He burning phase) using the equivalent evolutionary phases provided in the \boost{} models. 

The collision product is supplemented with the metals of the donor star. After the collision, the collision product will follow the newly assigned stellar track until another collision occurs or the star dies. 

To our knowledge, at the time of starting the simulation there were no stellar models exceeding $\sim500$--$600$ \mdot{} that would have also provided us with the chemical properties (e.g. Al, Na abundances) on the surface of the star throughout the stellar evolution. Our choice for the current study was thus to remain conservative and use the 500 \mdot{} \boost{} model at $Z=0.01$ \mdot{} to evolve any star that exceeds \mbox{500 \mdot}. This results in a lower limit for mass loss and the stellar radius. Recent extensions to the range of tables with initial stellar masses beyond \mbox{$\sim500$ \mdot{}} include the \textsc{PARSEC} tables that now reach \mbox{2000 \mdot{}} at low metallicities \citep{2025A&A...694A.193C}. For future applications we may consider updating our models to such tables. However, given that the \boost{} models are among the only ones that provide enough details of the chemical species related to GC MPs, we may alternatively explore the possibility of using tables such as provided by \citet{2025A&A...694A.193C} to build informed extrapolation techniques to extend the upper limit of 500 \mdot{} using the \boost{} models.

\subsubsection{Tidal disruption events}

A simple prescription for tidal disruption of massive stars by stellar black holes is also included. We detect a tidal disruption event (TDE) using the tidal radius given by
\begin{equation}
    r_\mathrm{tidal} = 1.3 r_\mathrm{*} \left(\frac{m_\mathrm{BH}+m_\mathrm{*}}{2m_\mathrm{*}}\right)^{1/3}
\end{equation}
where $r_\mathrm{*}$ and $m_\mathrm{*}$ are the current radius and mass of the star and $m_\mathrm{BH}$ is the mass of the black hole \citep{1992ApJ...385..604K}. Estimates of the exact amount of stellar debris accreted by the black hole depend on the internal structure of the star and the collision parameters and are still uncertain \citep{2013ApJ...767...25G}, ranging from a few per cent (e.g. \citealt{2016MNRAS.461..948M}) to unity \citep{2025PhRvD.111f3039B}. We follow the simple procedure often adopted in N-body and Monte Carlo simulations and deposit a fixed fraction (e.g. \citealt{2015MNRAS.454.3150G, 2023MNRAS.521.2930R}) of 50\% of the stellar mass into the black hole \citep{1988Natur.333..523R} and remove the other half from the simulation. The model can be further refined in the future to inject the returned mass back into the ISM. TDEs are extremely rare compared to stellar collisions in the current model since we only run the most intense starburst-phase for a short while. We will therefore leave exploration of TDEs as a source of chemical enrichment for future work.

\subsection{Initial conditions and simulation parameters}

The initial setup is based on a previous dwarf galaxy merger simulation presented in \citet{2024MNRAS.530..645L}. We re-simulate the starburst phase with the updated code version including collisional dynamics starting at a time when the starburst associated with the first passage is just starting to ramp up in the original simulation (simulation time 60 Myr in Fig. 4 in \citealt{2024MNRAS.530..645L}). Some small clusters have already formed by this time in the disks of the galaxies, however the main burst occurs in the overlap region between the galaxies and remains spatially separated from the pre-existing stellar disks for the duration of the burst. Those clusters that formed before the burst in the fully gravitationally softened simulation are here left out of the analysis.

The dwarf galaxies have virial and baryon masses of \mbox{$4\times 10^{10}$ \mdot{}} and \mbox{$6\times 10^{7}$ \mdot}, respectively, set up with the methods described in \citet{2005MNRAS.364.1105S}. The orbital configuration is the same as in \citet{2020ApJ...891....2L} but with reduced approach velocity (factor of 4) to achieve the starburst phase already after the first passage between the galaxies. The galaxies have initial gaseous and inert stellar disks with 66\% gas fraction and \mbox{0.73 kpc} scale lengths. The initial gas and stellar particle resolution is \mbox{4 \mdot{}} and the dark matter particle resolution is $6.8\times10^3$ \mdot. The initial metallicity is $\sim0.016$ \zdot. 

Gravitational interactions of dark matter, gas and pre-existing stars are softened, and the time integration is performed using the second order leapfrog integrator in \gadget. The gravitational softening length is set as \mbox{62 pc} for dark matter and \mbox{0.1 pc} for gas and pre-existing stars. Stars formed during the simulation have a gravitational softening length of \mbox{0.01 pc} if they are not within a \ketju{} integration region (see Section \ref{section:ketju}). Only the gravitational interactions of new stars and their remnants that occur within $r_\mathrm{KETJU}=0.03$ pc of any star with $m_i>8$ \mdot{} are solved using \ketju{} without gravitational softening.

We continue the starburst simulation for a total duration of \mbox{$\sim7.5$ Myr} until the largest stellar collision product exceeds \mbox{$\sim 1000$ \mdot}, i.e. when the most massive star considerably exceeds the upper limit of stellar evolution models (\mbox{$500$ \mdot}). By this time, the peak star formation rate of the starburst has passed, the most massive cluster has formed most of its stars, and the majority of the star clusters formed in the simulation have become gas free. 

\begin{figure*}
\includegraphics[width=\textwidth]{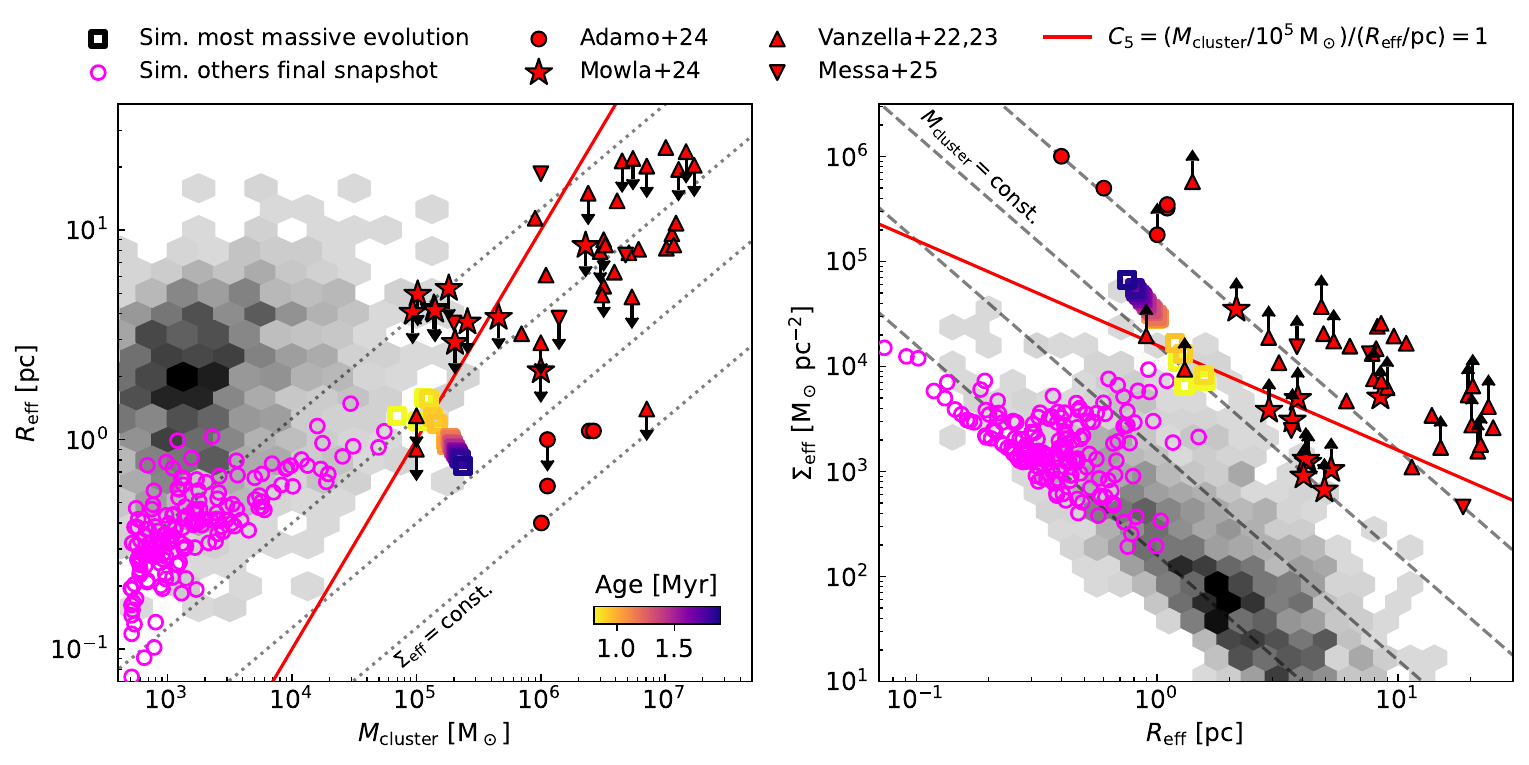}
\caption{The mass-size (\textit{left}) and size-surface density (\textit{right}) distributions of the simulated and observed young star clusters. The pink open circles show the general young star cluster population in the final snapshot. The squares indicate the evolution of the most massive cluster in \mbox{$\sim0.1$ Myr} steps starting when it had a bound mass equal to $25\%$ of its final mass (see Fig. \ref{fig:2dmaps}), coloured according to its mean stellar age from light to dark shade. The comparison data are observed young star clusters (\mbox{$<10$ Myr}) in the LEGUS survey (grey hexbins with the shade from light to dark indicating the number of clusters ranging from 1 to 38 clusters on the left and 1 to 28 clusters on the right; \citealt{2021MNRAS.508.5935B}) and various surveys of strongly lensed clusters at redshifts $z\gtrsim2.4$ (filled symbols; \citealt{2022A&A...659A...2V, 2022ApJ...940L..53V, 2023ApJ...945...53V, 2024Natur.632..513A, 2024Natur.636..332M, 2025A&A...694A..59M}), for which we have used the reported half-light radii to compute the effective surface densities. The solid line shows the compactness parameter $C_5=(M_\mathrm{cluster}/10^5\, \mathrm{M}_\odot)/(R_\mathrm{eff}/\mathrm{pc})=1$ \citep{2016A&A...587A..53K}. The dotted lines on the left show sizes corresponding to a constant surface density for masses in the range $10^3$--\mbox{$10^6$ \mdot{} pc$^{-2}$} from left to right and the dashed lines on the right show the surface densities for a constant mass in the range $10^3$--\mbox{$10^6$ \mdot{}} from left to right, in steps of 1 dex.
\label{fig:mass_size}}
\end{figure*}

\subsection{Cluster selection}

We identify gravitationally bound star clusters using the \textsc{friends-of-friends} (\textsc{fof}) and \textsc{subfind} \citep{2001MNRAS.328..726S, 2009MNRAS.399..497D} algorithms in \gadget. The linking length of \textsc{fof} is set as \mbox{0.2 pc} to prevent the algorithm from including extended sparse structures outside of the concentrated clusters of stars. A bound structure has to have at least 50 members after the unbinding step of \textsc{subfind} to be considered a cluster. After the selection of bound clusters we also include any stars within the half-mass radius regardless of their velocity, to catch stars that are in binaries or escaping but have not yet exited the central region of the cluster. These stars are included in the mass, light and density calculations but not in the computations of the velocity distribution.

\subsection{Radiative transfer post-processing}

We produce photometric images of the most massive cluster using the dusty Monte Carlo radiative transfer code \skirt{} 9 \citep{2020A&C....3100381C}. We follow the methods for setting up the input radiation sources and the construction of the dust grid described in \citet{2022MNRAS.514.4560L, 2023MNRAS.522.3092L, 2025MNRAS.537..956P}. Briefly, all stars more massive than \mbox{0.8 \mdot{}} that have formed during the simulation are used as radiation sources with spectra given by atmosphere models of \citet{2003IAUS..210P.A20C}. As with stellar evolution, stars beyond $500$ \mdot{} are treated as stars with initial mass of \mbox{$500$ \mdot{}} for consistency. The dust distribution is computed as a (60 pc)$^3$ octree grid constructed according to the SPH particle data and a fixed $\sim80\%$ dust-to-metals mass ratio. The grid is allowed to refine at most down to $\sim 0.01$ pc. Only gas below a temperature of \mbox{8000 K} is considered to contain dust. Photometric images are then made at $\sim0.04$ arcsec per pixel resolution that is close to the pixel resolution of \textit{Hubble Space Telescope} (HST) and \textit{James Webb Space Telescope} (JWST), placing the cluster at 50 kpc distance that is similar to the distance of R136 in LMC. We use three standard broadband Johnson-Cousins filters B, V and I built in \skirt{}. We apply a Gaussian point spread function with a full width at half maximum of 2 pixels.

\section{Results}\label{section:results}

\subsection{Properties of young star clusters}

Approximately 850 star clusters that fill our selection criteria form during the $\sim 7.5$ Myr of starburst. Star clusters above \mbox{$\sim 300$ \mdot{}} exhibit a cluster mass function that has a power-law index between $-1.9$ and $-2$ throughout the starburst, fit using 20 bins with equal number of 40--50 clusters per bin. The slope of the mass function agrees with observed mass functions and luminosity functions across environments from low-metallicity galaxies such as the Magellanic Clouds to starburst systems \citep{1996ApJ...471..816E, 1997ApJ...480..235E, 1999ApJ...527L..81Z, 2003AJ....126.1836H, 2009A&A...494..539L, 2013MNRAS.431..554R}.

By the final snapshot, the majority of the clusters are already gas-free while all clusters have mean stellar ages less than \mbox{4.3 Myr}. The most massive cluster at a mean stellar age of \mbox{1.9 Myr} still has a gas-embedded core that is forming stars. While the tail end of the starburst is still ongoing, we limit the analysis only until this epoch. The majority of the stellar mass has already formed, and the main stellar collision product (Section \ref{section:VMSs}) in the most massive cluster has not yet reached a mass significantly above \mbox{1000 \mdot}, i.e. more than a factor of two beyond the upper stellar mass limit adopted in our VMS models (see Section \ref{section:collisions}).
Fig. \ref{fig:mass_size} shows the projected half-mass radii ($R_\mathrm{eff}$) and the corresponding surface densities $\Sigma_\mathrm{eff}=M_\mathrm{cluster}/(2\pi R_\mathrm{eff}^2)$ in the $x$--$y$ plane for the simulated gravitationally bound star clusters in the final snapshot. The results are compared to masses and half-light radii of young ($<10$ Myr) star clusters in the LEGUS survey \citep{2021MNRAS.508.5935B} and of resolved or marginally resolved star clusters measured through gravitational lensing at redshifts $z\gtrsim 2.4$ (\citealt{2022ApJ...940L..53V, 2022A&A...659A...2V, 2023ApJ...945...53V, 2024Natur.632..513A, 2024Natur.636..332M, 2025A&A...694A..59M}). The surface densities in the observed clusters have been computed using the reported half-light radii.

The majority of the simulated star clusters show compact sizes of $R_\mathrm{eff}\sim0.1$--$1$ pc, though only the most massive cluster reaches surface densities of close to $10^5$ \mdot{} pc$^{-2}$. The range of initial sizes is in general agreement with compact sizes found in our previous studies in \citet[][resolution 4 \mdot]{2020ApJ...891....2L}, \citet[][fully sampled IMF, softened gravity]{2024MNRAS.530..645L} and \citet[][fully sampled IMF, incl. collisional dynamics with \ketju]{2025MNRAS.538.2129L} for young clusters. Cloud-scale simulations of massive star clusters often also report compact sizes, e.g. in \citet{2018MNRAS.478.4142T, 2024A&A...690A..94P}. Clusters in the new simulation that have initial masses of a few $10^2$--$10^5$ \mdot{} are more extended on average by a factor of 1.5--2 compared to clusters of similar mass and age in the previous setup of \citet{2024MNRAS.530..645L} without collisional dynamics near massive stars. This difference is also seen as a slightly steeper mass-size relation slope of 0.32 compared to 0.26 fit to the clusters in \citet{2024MNRAS.530..645L}. Similar to the results of \citet{2025MNRAS.538.2129L} for low-mass clusters, the fully exposed clusters undergo rapid size evolution due to internal gravitational dynamics already during the very first Myrs. 

Our clusters have sizes that correspond to the most compact and dense young star clusters in local observations. Observationally it remains unclear whether very young clusters in general begin their lives with smaller sizes and higher densities compared to evolved clusters as suggested by theoretical models (e.g. \citealt{Marks2012, 2017A&A...597A..28B}) and our simulations. The larger sizes and lower surface densities of present-day GCs (\citealt{2005ApJS..161..304M}) compared to their high-redshift counterparts imply some evolution at least over extended periods of time. The size-mass relation of large samples of low-redshift star clusters e.g. in the LEGUS survey (e.g. \citealt{2017ApJ...841..131A}, \citealt{2021MNRAS.508.5935B}; see also \citealt{2019ARA&A..57..227K} for a discussion) instead show on average very little evolution with cluster age. Interestingly, \citet{2017ApJ...841...92R} for instance verified that selection effects due to point-source removal should not be the cause for missing centrally concentrated bright massive clusters in the local Universe. On the other hand, young clusters may reach peak densities while they are still embedded, remaining mostly invisible to observations in visual wavelengths \citep{2023ApJ...946....1C}. High-resolution infrared observations e.g. with the JWST may bring more light to this aspect in near future \citep{2023ApJ...944L..14W, 2024ApJ...973L..55L}.

In support of the latter argument, our compact clusters in Fig. \ref{fig:mass_size} are very young and the majority have just become fully gas-free. As was shown in \citet{2025MNRAS.538.2129L}, the star clusters first become more compact while they are still embedded and forming stars. The most massive clusters in the present study form hierarchically through accretion of gaseous and stellar sub-clumps, and their internal structure will undergo violent relaxation \citep{1967MNRAS.136..101L}. The time evolution of the most massive star cluster is shown here in Fig. \ref{fig:mass_size} as a specific example, starting when the most massive progenitor cluster had a bound mass of 25\% of the total mass of the final cluster. The contraction of the cluster can be seen in the time-sequence in Fig. \ref{fig:mass_size}. Similar contraction was also shown in cloud-scale simulations of massive star cluster formation by \citet{2024A&A...690A..94P}.

Finally, we also show the compactness parameter $C_5=(M_\mathrm{cluster}/10^5\, \mathrm{M}_\odot)/(R_\mathrm{eff}/\mathrm{pc})$ with a value of $C_5=1$ in Fig. \ref{fig:mass_size}. This was estimated in \citet{2016A&A...587A..53K} to be the approximate limit above which gas expulsion is no longer possible due to failure of stellar feedback (see also e.g. \citealt{2017MNRAS.465.1375S}, \citealt{2019MNRAS.483.5548G} and \citealt{2018ARA&A..56...83B}). Our most massive star cluster begins its life close to this relation, and progresses toward larger values of $C_5$ as it grows in mass and contracts. All this happens while VMSs are forming and growing in its core and releasing large amounts of pre-SN feedback. The $C_5=1$ line divides the mass-size and size-surface density planes into two regimes, where our simulated most massive cluster falls in the same regime with the high-redshift dense clusters. Our lower mass simulated clusters and the local observed young star clusters populate the $C_5<1$ regime. As suggested in \citet{2016A&A...587A..53K} and supported by the simulations in \citet{2024MNRAS.530..645L}, $C_5>1$ may well be the dividing line above which self-enrichment and MPs start occurring. Overall, the nitrogen enhancements observed in some high-redshift compact galaxies (e.g. \citealt{2023A&A...673L...7C, 2023ApJ...959..100I, 2024MNRAS.529.3301T, 2024A&A...681A..30M}), the high compactness of the corresponding high-redshift clusters and their high stellar masses are in line with the approximate initial GC mass limit of \mbox{$\gtrsim 10^5$ \mdot{}} for Local Group GCs to host MPs \citep{2019A&ARv..27....8G}.

\subsubsection{Stellar collisions and two TDEs}
In total $\sim12500$ stars more massive than $8$ \mdot{} form throughout the starburst. 69 are more massive than $100$ \mdot. When stellar collisions occur in our simulation, only one star grows per cluster after segregating into the cluster core, similar to results in the literature \citep{1999A&A...348..117P, Baumgardt2011,2013MNRAS.430.1018F,2024MNRAS.531.3770R}. In two cases the merger product of two stars below the limit of 100 \mdot{} exceeds the VMS limit. The rest of the events that grow VMSs start with a star originally above 100 \mdot. The most massive star cluster hosts $>2000$ stars more massive than $8$ \mdot, up to $\sim18$ of which are more massive than $100$ \mdot{} and present in the cluster at the same time. 12 remain in the cluster in the final snapshot. 13 of the VMSs form after the main progenitor has reached 25\% of its final mass and they form within $\sim1$ pc of the cluster centre. The rest arrive via sub-cluster mergers. By the last snapshot, $78$ stars  in the entire galaxy more massive than $40$ \mdot{} have collapsed as black holes, out of which $8$ remain in the most massive cluster. 20 stars have exploded as pair-instability supernovae.

For comparison, R136 includes tens of stars that are more massive than a few tens of \mdot. Of these $\sim10$ may have started initially with masses of more than \mbox{100 \mdot{}} \citep{2022A&A...663A..36B}, the most massive estimated to have a current mass of $\sim220$ \mdot{} and an initial mass of $\gtrsim250$ \mdot. The most massive stars in R136 are currently located in the central \mbox{$\sim 0.5$ pc} of the cluster. The number of VMSs in our most massive cluster is thus approximately twice as many as in the R136, consistent with the higher total cluster mass. 

We track the collisions of stars throughout the simulation for pairs that include a star or a black hole that had an initial stellar mass of at least $9$ \mdot. During the $\sim7$ Myr of starburst there are in total $90$ collisions throughout the simulation, out of which $16$ involve two components both of which are more massive than \mbox{$40$ \mdot}. Almost $30\%$ of the collisions are between fairly equal mass objects (mass ratio $>1/3$). We note that compared to the hierarchical direct N-body simulations of \citet{2024MNRAS.531.3770R}, we do not see the growth of VMSs through collisions in the in-falling sub-clusters during the assembly of the most massive cluster. This may be due to somewhat delayed mass segregation and sub-cluster core collapse in our collisional dynamics implementation, compared to full direct N-body (see Fig. 1 in \citealt{2025MNRAS.538.2129L}). The core collapse time scale, for instance, affects when VMS growth can occur in the sub-clusters \citep{2013MNRAS.430.1018F}. Optionally the spread in stellar ages in each cluster due to the star formation history may delay the birth of VMSs and their expansion into the giant phase compared to the uniform age initial conditions adopted in N-body models. This shifts the collisional growth of VMSs to the later, more centrally concentrated stage of cluster assembly.

There are two (micro-)TDEs: one between a \mbox{58 \mdot{}} black hole and a \mbox{125 \mdot{}} VMS (age \mbox{$\sim 2.3$ Myr} and initial mass \mbox{132 \mdot}), and the other between a \mbox{160 \mdot{}} black hole and a \mbox{74 \mdot{}} VMS (age \mbox{$\sim 2.9$ Myr} and initial mass \mbox{89 \mdot}). In the first TDE the VMS had recently reached its supergiant phase. In the second one the star was just about to collapse as a black hole. The collisions occurred not in the most massive cluster but in the cores of a $2\times10^4$ \mdot{} and a $3\times10^4$ \mdot{} cluster that had evolved to a mean stellar age of 2.7 Myr and 2.9 Myr, respectively. Black hole mergers do not occur in the simulation essentially because the post-Newtonian terms in \ketju{} were disabled for the run.

There are several factors that affect the collision rate obtained in the simulation. Because close gravitational interactions are only possible in systems that are integrated with \ketju{} and mergers are only allowed to happen for pairs that contain a $m_i>8$ \mdot{} star, the collision rate found here is a lower limit. On the other hand, the self-consistent star formation history that we obtain for each cluster adds complexity to the stellar evolution and collisions compared to pure N-body simulations that often assume uniform age clusters. The stars for instance evolve to their giant stage later if they form later in the cluster formation process. This will naturally reduce the collision rate during the assembly process due to the reduced collision cross-section. Meanwhile, the most massive stars mainly form in the central regions of the clusters and mass segregation of the most massive collision partners can thus happen on a shorter time scale. We omit the initialisation of primordial binaries, which would additionally enhance the collision rate \citep{2004MNRAS.352....1F, 2025MNRAS.542L..78R}.

\begin{figure*}
\includegraphics[width=\textwidth]{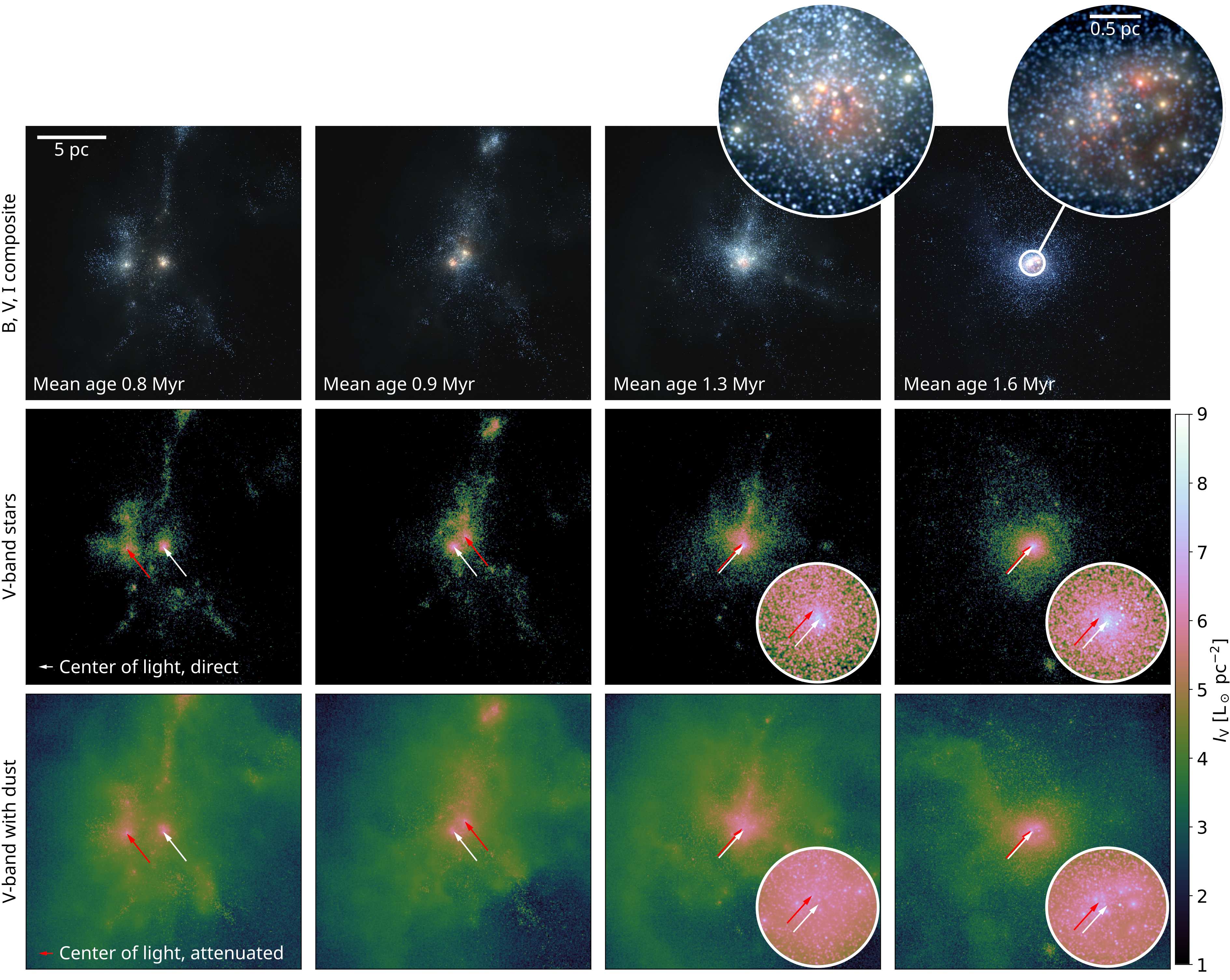}
\caption{\textit{Top}: Colour composite images of the central \mbox{20 pc} at four epochs along the formation of the most massive cluster (see text) in B (blue), V (green) and I (red) Johnson-Cousins broad-bands. The image resolution corresponds to that of the HST at the distance of LMC (\mbox{$\sim 50$ kpc}, \mbox{0.01 pc} per pixel) degraded with a \mbox{$\sim2$ pix} Gaussian point spread function. The insets show the central \mbox{2 pc} in an adjusted value range. 
\textit{Middle}: V-band surface brightness maps of the stellar emission without dust attenuation. The white arrows show the centres of light (see text for details). The brightest spot close to the centre of the rightmost panel corresponds to the most massive star (see Section \ref{section:VMSs}) whose radiation properties are modelled as an initially 500 \mdot{} star at an age of 1.2 Myr. This star, among others, is not visible in the top and bottom rows due to extinction. The insets in the middle and bottom rows show the central 2 pc in the same value range.
\textit{Bottom}: V-band surface brightness maps with attenuation, scattering and dust emission. The red arrows indicate the centres of light of the attenuated images measured similarly to the unattenuated one to highlight the offset in the direct and attenuated light.
\label{fig:2dmaps}}
\end{figure*}

\subsubsection{The most massive star}\label{section:VMSs}
In the most massive cluster, one star grows to a peak mass of \mbox{1046 \mdot} through repeated collisions with other massive stars. It began its life as a \mbox{150 \mdot{}} star, and consequently merged with stars of \mbox{115 \mdot}, \mbox{170 \mdot}, \mbox{191 \mdot}, \mbox{232 \mdot}, \mbox{153 \mdot{}} and \mbox{44 \mdot{}} during a period of 1 Myr. The collisions are spaced by 0.1 Myr, 0.15 Myr, 0.4 Myr, 0.3 Myr and 4.4 kyr. The most massive star also merged with 13 low-mass stars ($<0.4$ \mdot), increasingly frequently during the final 0.2 Myr as it grew to the supergiant phase. We limit the analysis to this epoch, however the star would still have the potential to collide with a variety of other massive stars, including the most massive normal (i.e. not collisionally formed) VMS of 330 \mdot. This star is at a distance of only \mbox{0.05 pc} from the cluster core and the most massive star in the final snapshot. 

The final VMS mass that far exceeds the upper stellar mass limit of the adopted IMF, and its growth being dominated by collisions with other VMSs, are consistent with hierarchical direct N-body models of star cluster assembly \citep{2024MNRAS.531.3770R}. All of the individual collision partners are within the range of estimated initial masses of the most massive stars in R136, therefore the physical scenario for the VMS growth through collisions is very well motivated based on observations. Why no such massive stars that could be collision products are observed in R136 may be due to its too low central/effective density, or its higher metallicity (factor of tens) that allows stronger mass-loss and smaller stellar radii at evolved stages.

Our resulting growth of the most massive star is less extreme than what is found in idealised cloud-scale hydrodynamical simulations (e.g. \citealt{2023MNRAS.521.3553R, 2024Sci...384.1488F, 2025MNRAS.539.2561C}). Such simulations of dense star-forming clouds find VMSs in excess of $10^3$--$10^4$ \mdot{} forming through collisions and/or gas accretion. Future work with improved treatment for the evolution of the most massive stars will help address whether this difference is a result of the galactic environment and more complex cloud structure in our model, the stellar evolution models, the adopted IMF, or perhaps something else.

Though we have refrained from making assumptions about the detailed evolution of stars beyond \mbox{$500$ \mdot}, an order of magnitude estimate for the final mass of the most massive star and its black hole remnant can be made based on simplified arguments about the mass and metallicity dependent wind mass loss e.g. by \citet{2018A&A...615A.119V} and \citet{2023MNRAS.524.1529S}. The most massive star undergoes its last major collision at an age of $\sim 1.2$ Myr, and it can be expected to still live for one million years. A $\sim1000$ \mdot{} star at $\sim 0.01$ \zdot{} would initially lose mass in optically thick winds at a rate between $\sim 10^{-4.3}$--$ 10^{-3.7}$ \mdot{} yr$^{-1}$ depending on the rate estimate. The star would lose in total 
between $\sim25$--$160$ \mdot{} during the final \mbox{1 Myr} of its life, resulting in a black hole with a mass of at least \mbox{$\sim840$ \mdot}. Even if the star would be rotating fast as a consequence of its merger history and therefore lose mass at a factor of 2--3 faster rate as argued by \citet{2025MNRAS.tmp.1257G}, it would still result in a final black hole mass of at least $640$ \mdot. Additionally, if the stellar collisions were implemented with a prescription for mass loss, we can approximate the star to end up with 10\% lower maximum mass, $\sim900$ \mdot, after the last major  collision. This would further decrease the resulting black hole mass to a minimum of \mbox{600 \mdot{}} (rotating VMS with \citealt{2023MNRAS.524.1529S} mass loss rate) or 770 \mdot{} (without rotation). The result would therefore be an intermediate mass black hole in a proto-globular cluster-like system. Whether such a black hole would be retained in the cluster or the galaxy over several Gyrs would require a more detailed modelling of black hole--black hole interactions and mergers in the simulation (see previous studies e.g. by \citealt{2014ApJ...782...97A}, \citealt{2021A&A...652A..54A}, \citealt{2024MNRAS.531.3770R}).

\begin{figure*}
\includegraphics[width=\textwidth]{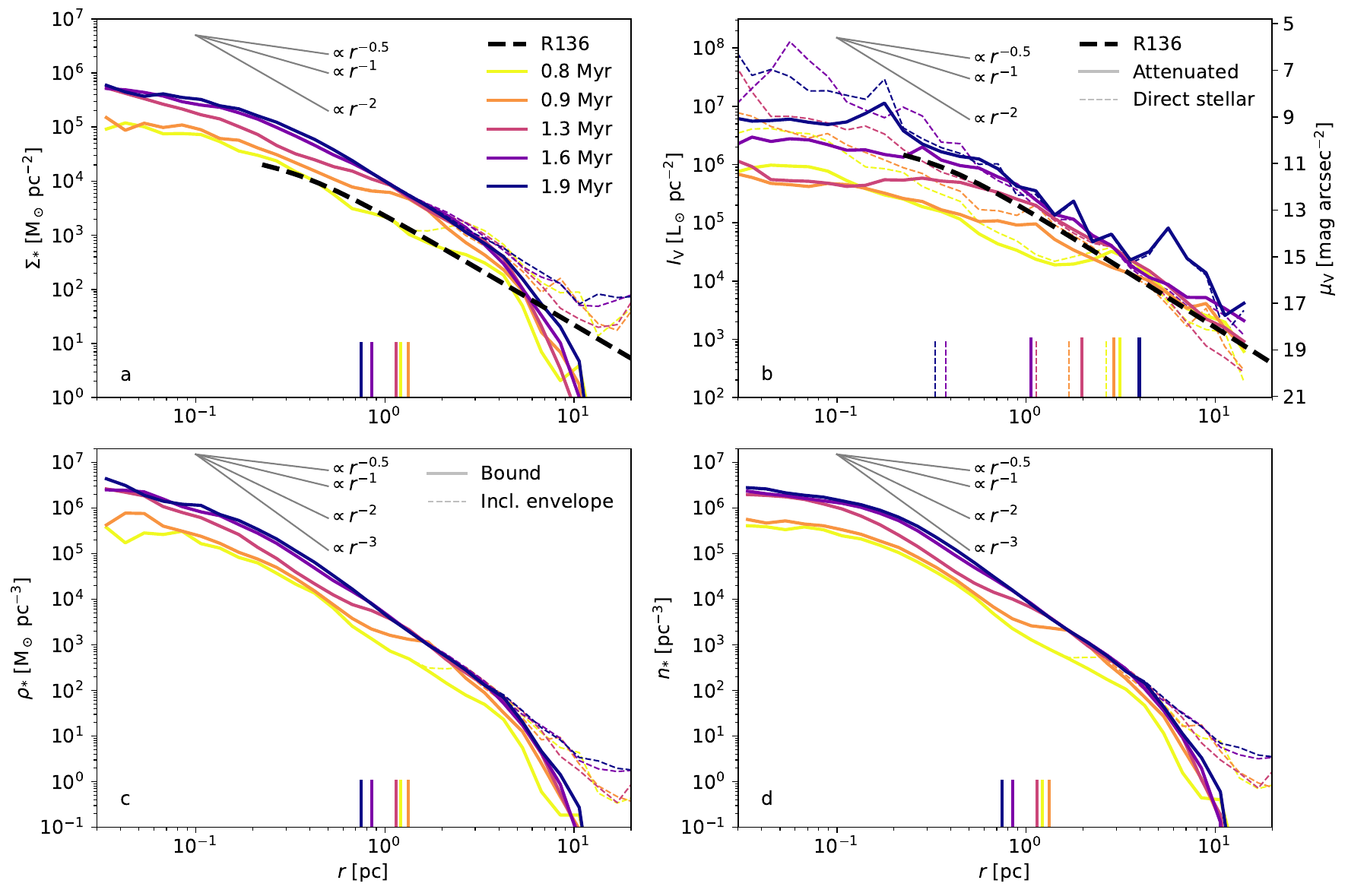}
\caption{The radial projected profiles of stellar mass surface density (\textit{a}), V-band surface brightness (\textit{b}), 3D mass density (\textit{c}) and stellar number density (\textit{d}) of the most massive cluster at the four epochs shown in Fig. \ref{fig:2dmaps} and in the last snapshot when the cluster has a mean stellar age of \mbox{1.9 Myr}. The solid lines in panels \textit{a}, \textit{c} and \textit{d} indicate the gravitationally bound stellar component and the dashed lines show the total stellar mass in the region. In panel \textit{b} the solid and dashed lines show the attenuated and the unattenuated surface brightness profiles, respectively. The power-law slopes of $-0.5$, $-1$, $-2$ and $-3$ are indicated in the panels. The best-fit surface density and brightness profiles of R136 from \citet{2005ApJS..161..304M} are shown in black in panels \textit{a} and \textit{b}, and the right-side y-axis in panel \textit{b} shows the surface brightness scale in mag arcsec$^{-2}$. The vertical bars at the bottom of the panels show the projected half-mass radii of the bound component (panels \textit{a}, \textit{c} and \textit{d}) and the projected half-light radii of the attenuated and unattenuated light profiles (\textit{b}) with respective linestyles.
\label{fig:densities}}
\end{figure*}

\subsection{Properties of the most massive cluster}\label{section:MMC}

Fig. \ref{fig:2dmaps} shows a time sequence of the accumulation of mass in the most massive star cluster. The panels show colour composites and surface brightness maps of the direct and attenuated stellar light at four epochs. The epochs shown correspond to when the gravitationally bound main progenitor is 25\%, 50\%, 75\% and $\sim90\%$ of the final mass ($\sim 2.2\times 10^5$ \mdot) of the most massive cluster. The mean stellar age of the main progenitor cluster is \mbox{$0.8$ Myr}, \mbox{$0.9$ Myr}, \mbox{$1.3$ Myr} and \mbox{$1.6$ Myr} in snapshots that are spaced by \mbox{$0.2$ Myr}, \mbox{$0.4$ Myr} and \mbox{$1.1$ Myr}, the last being $\sim0.4$ Myr before the final snapshot. The visual appearance of the most massive cluster in the final snapshot does not show additional features except for having less gas and a more compact central concentration of massive stars, therefore we only show the final surface density profile later in Section \ref{section:profiles}. Throughout the assembly, the mean gas surface density within \mbox{1 pc} is $\sim 3$--5 \mdot{} pc$^{-2}$, i.e. at least an order of magnitude less than that of the stellar component.

The first two epochs show the phase of cluster assembly where stellar clumps are concurrently coalescing and forming stars. The stellar structure at the top of the panels in the two left columns is falling toward the cluster. The centre of light indicated in the V-band images is measured with a 2D shrinking sphere of minimum radius of 10 pixels, masking out all pixels brighter than $10^5$ \ldot{} (i.e. massive stars) before applying the point spread function. This prevents the algorithm from centring on the brightest stars. 

The stellar light of the main cluster is attenuated and the other massive clump in the images remains the brightest cluster in V-band for the duration of this coalescence, irrespective of the masking applied in the centring. The V-band attenuation $A_V$ is $\sim 0.9$ mag at these epochs, computed as the AB-magnitude difference between the unattenuated and attenuated images within a radius of 1 pc of the centre of light of the unattenuated image. This is a moderate value for observed star clusters  and not as extreme as the most attenuated clusters observed in some dwarf galaxies (e.g. \citealt{2023ApJ...946....1C}). The reddening $E(B-V)$ within the same 1 pc region is 0.7 and 0.5 in the two first epochs, similar to what is typically found for young clusters in dwarf galaxies or starbursting systems ($E(B-V)$ up to 0.7 e.g. in \citealt{2020MNRAS.499.3267A, 2020ApJ...889..154W}).

The last two epochs in Fig. \ref{fig:2dmaps} show centrally concentrated star formation. The central 1 pc has a star-forming disky gas component until the last snapshot. The gas disk is being fed along a gaseous filament that extends toward the cluster from north-west in the rightmost panels. $A_V$ peaks at $\sim 1.4$ during the last epochs, and $E(B-V)$ drops to 0.3--0.4.

\subsubsection{Density, surface density and surface brightness profiles}\label{section:profiles}

In Fig. \ref{fig:densities} we show the build up of the radial density, surface density and V-band light profile of the most massive cluster at epochs corresponding to Fig. \ref{fig:2dmaps} as well as in the final snapshot. The particle data is centred with a shrinking sphere ignoring all stars more massive than 50 \mdot, similar to the brightness profiles. 

The density and brightness profiles of the most massive cluster are consistent with power-law shapes especially in the outer parts and at later epochs when the stellar distribution is less clumpy (see Fig. \ref{fig:2dmaps}). The infalling clusters in our simulation can be seen as an increase in density at intermediate radii ($1$--$3$ pc) at earlier epochs. The outer radial power-law slope in surface density is $-2.2$, as expected from models of hierarchical sub-clusters assembly that predict slopes that approach $-2$ \citep{2018MNRAS.477.1903S, 2018MNRAS.481..688G, 2024MNRAS.531.3770R}. The outer power-law  has a respective density slope of slightly shallower than $-3$ and continues well into the sub-parsec regime. The final central 3D mass density, 3D number density and surface mass density (inner $\sim1000$ stars) are a factor of $\sim190$, 10 and 10 times higher than the half-mass values shown in Fig. \ref{fig:mass_size}, respectively. The unattenuated central surface brightness profile likewise has a steep inner slope, dominated by the most massive stars in the cluster core. Based on these simulated brightness and density profiles, we therefore infer that the central surface densities $\Sigma_0$ of the observed high-redshift clusters may be at least an order of magnitude higher than the effective values $\Sigma_\mathrm{eff}$, i.e. $\Sigma_0>10^6$ \mdot{} pc$^{-2}$. This may be even before the clusters undergo core collapse, based on our simulated cluster that has only just formed. The fact that the cluster is still forming stars in a central disk is consistent with the idea that the central stellar density of \mbox{$>5\times 10^5$ \mdot{} pc$^{-2}$} is high enough to prevent pre-SN stellar feedback from expelling the gas \citep{2019MNRAS.483.5548G}. 

The number density profile shows a shallow central profile that is an expected feature in N-body simulations of repeated star cluster mergers \citep{2012ApJ...750..111A, 2024MNRAS.531.3770R}. The relaxation time of the core is approximately one Myr, therefore star formation and relaxation occur simultaneously, together with mass segregation. The spatially resolved scale of the stellar component in the simulation is a few \mbox{$0.01$ pc} or less, however we note that the star sampling routine places stars in a Gaussian kernel of \mbox{$\sigma=0.1$ pc} that impacts the central-most density profile of the youngest stars. We therefore refrain from making quantitative statements regarding the density profile in the core region. 

We also show a comparison to the best-fit surface density and V-band surface brightness profiles of the young massive cluster R136. The inner radius is limited to the range of data typically used in the fits such as provided by \citet{2003MNRAS.338...85M} and \citet{2005ApJS..161..304M}. As with young clusters in general \citep{2005ApJS..161..304M}, the radial surface brightness profile of R136 is well fit with an Elson-Fall-Freeman profile \citep{1987ApJ...323...54E}. The profile of R136 has an outer power-law slope of the order of $-2$ within the central \mbox{$\sim 5$ pc} \citep{2003MNRAS.338...85M, 2005ApJS..161..304M} and possibly a small central core of \mbox{$\sim0.3$ pc}. The central R136 cluster is less massive than our simulated cluster by a factor of a few, nonetheless its surface density and light profiles are qualitatively very similar to our simulated cluster. The central mass density of R136 has been estimated to be between $10^4$--$10^7$ \mdot{} pc$^{-3}$ \citep{2013A&A...552A..94S}, depending on the assumed size of the central core. Such values are supported by our hydrodynamical results throughout the cluster assembly.

\begin{figure*}
\includegraphics[width=\textwidth]{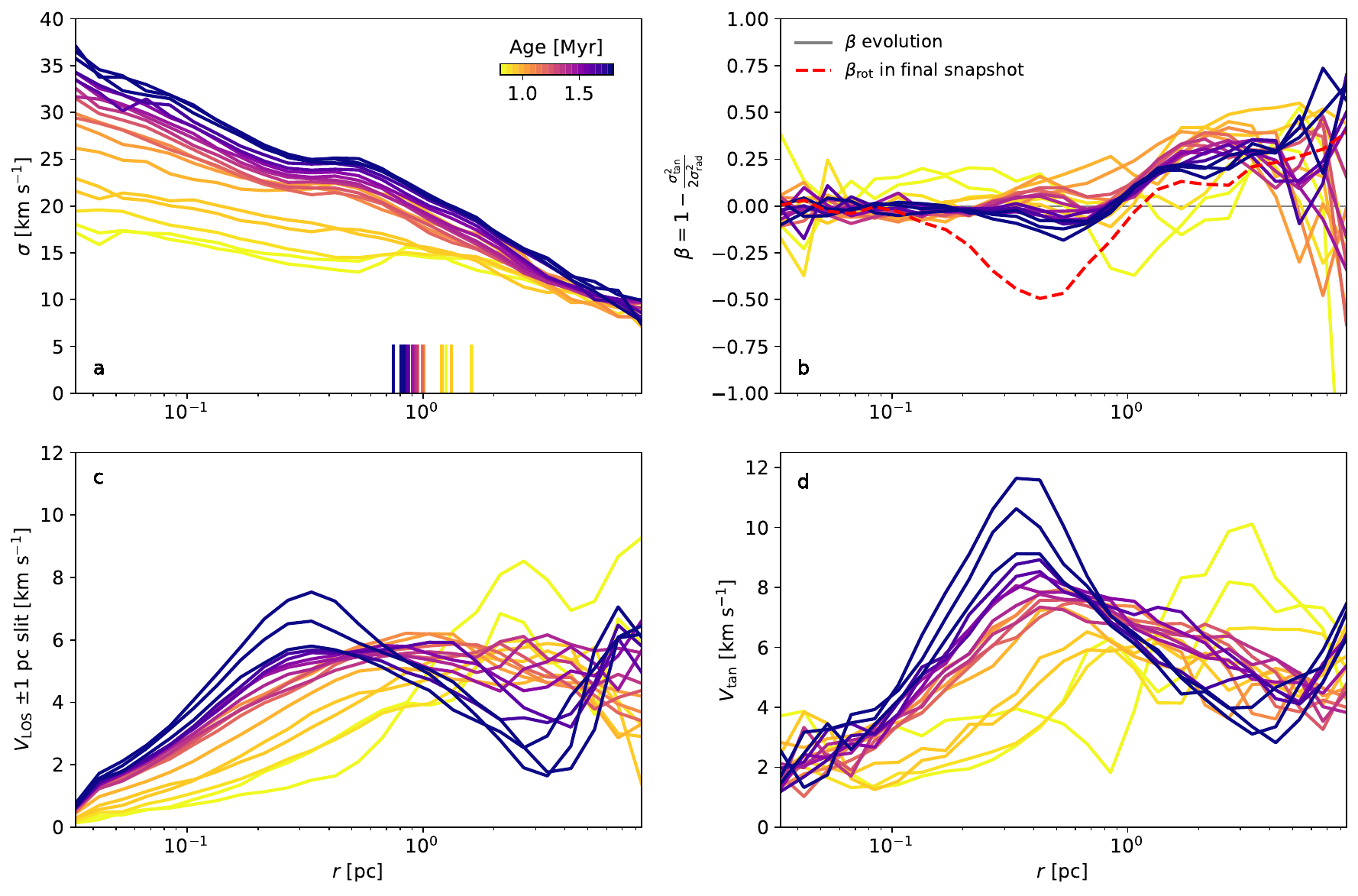}
\caption{The radial velocity distribution of the most massive cluster in \mbox{0.1 Myr} steps. The lines are coloured with the mean stellar age of the cluster, increasing from light to dark. The panels show the 3D velocity dispersion (\textit{a}), the velocity anisotropy (\textit{b}), the rotation velocity along a $\pm 1$ pc slit along an inclined plane of rotation (\textit{c}) and the tangential velocity (\textit{d}). Each line is a mean of 100 bootstrapped samples oriented in a random inclination. The dashed line in panel \textit{b} shows the final anisotropy profile without the removal of the bulk motion, $\beta_\mathrm{rot}$. The vertical bars in panel \textit{a} show the projected $R_\mathrm{eff}$.
\label{fig:vels}}
\end{figure*}

\subsubsection{Velocity distribution}

Next we inspect the velocity dispersion and velocity anisotropy profiles, as well as whether the cluster rotates. We compute the radial and tangential velocity distributions in spherical coordinates $v_\mathrm{rad}$ and $v_\mathrm{tan}=\sqrt{v_\theta^2 + v_\phi^2}$ (where $v_\theta, v_\phi$ are the angular velocity components) as well as using the line of sight (LOS) velocity in a $\pm1$ pc slit perpendicular to the cluster rotation axis, $V_\mathrm{LOS}$. The slit is placed according to the angular momentum vector within the 3D half-mass radius. The velocities are binned radially in \mbox{0.1 dex} bins and the innermost bin includes all stars within that radius. The corresponding velocity dispersions $\sigma_\mathrm{rad}, \sigma_\theta, \sigma_\phi$, and $\sigma_\mathrm{tan}=\sqrt{\sigma_\theta^2 + \sigma_\phi^2}$ and total $\sigma=\sqrt{\sigma_\mathrm{rad}^2+\sigma_\theta^2 + \sigma_\phi^2}$ are computed using the standard deviation of each velocity component in each bin. The cartesian coordinates are centred at the cluster centre of mass. Here we also mask out stars that have velocities greater than \mbox{$\sim70$ km s$^{-1}$}, in excess of the bulk velocity distribution in the cluster, to remove stars that are either escaping or in close binaries. Standard deviations for the profiles are computed by bootstrapping the stars 100 times, each time in a random orientation to include the effect of inclination on $V_\mathrm{LOS}$. 
We use the velocity anisotropy parameter $\beta$ to describe the ratio in tangential and radial dispersion \citep{2008gady.book.....B} as 
\begin{equation}
    \beta=1-\frac{\sigma_\mathrm{tan}^2}{2\sigma_\mathrm{rad}^2}. 
\end{equation}

Fig. \ref{fig:vels} shows the evolution of the velocity distribution in the most massive cluster with time. The increase of the velocity dispersion toward the centre steepens with time as the central density of the cluster increases (see Fig. \ref{fig:densities}). The peak central value is $\sigma\sim 37$ km s$^{-1}$. Once the cluster settles with one dominant core (as opposed to several merging sub-clumps), its peak rotation velocity shifts inwards. By the final snapshot the peak value reaches \mbox{$V_\mathrm{LOS}\sim7$ km s$^{-1}$} and \mbox{$v_\mathrm{tan}\sim12$ km s$^{-1}$} at a radius of $\sim 0.3$--\mbox{$0.4$ pc}. For comparison, GCs have similar or lower rotation velocities measured both in 1D and 3D \citep{2018MNRAS.473.5591K, 2018MNRAS.481.2125B, 2025A&A...694A.184L}, consistent with their dynamically more evolved state. R136 also rotates at a rate of a few km s$^{-1}$ \citep{2012A&A...545L...1H}, which is slower than our simulated cluster and consistent with its lower mass and previous results on star cluster dynamics in the \griffin{} project \citep{2020ApJ...904...71L}. 

The velocity distribution remains fairly isotropic within the inner \mbox{1 pc}, while the outer parts develop with a radially biased velocity distribution. Qualitatively similar results were found by \citet{2025ApJ...984...75K} in cloud-scale simulations of hierarchical assembly, where the outer regions of the final cluster show radial anisotropy. The stellar population giving rise to the rotation signal and to the anisotropy in our cluster is discussed in more detail in Sections \ref{section:populations} and \ref{section:kinematics}.

As described in \citet{2025A&A...694A.184L}, the anisotropy profile can be computed using the standard deviations or the root mean square velocities. The former gives the intrinsic anisotropy, while the latter includes the bulk velocity component in each radial bin. That is, anisotropy including bulk motion $\beta_\mathrm{rot}$ is computed as $\sigma^2=\frac{1}{N} \sum_{i=0}^{N}{v_i}^2$ instead of $\sigma^2=\frac{1}{N} \sum_{i=1}^{N}{(v_i-\Bar{v})^2}$ where $\Bar{v}$ is the mean velocity per bin. The bulk motion mostly composes of rotation if the cluster is rotating and not expanding or contracting. The final $\beta_\mathrm{rot}$ of the simulated cluster is shown in Fig. \ref{fig:vels}. The bulk rotation in the inner radii is reflected in a tangentially biased $\beta_\mathrm{rot}$. The radially dominated $\beta_\mathrm{rot}$ at outer radii is reduced compared to $\beta$, however the anisotropy remains mildly positive even when the bulk motion in each component is included. 

The rotational support $(V/\sigma)_\mathrm{tot}$, computed as the ratio between peak rotation velocity within $3 R_\mathrm{eff}$ and the central velocity dispersion, is $\sim0.3$--$0.5$ regardless of using the LOS velocity or the tangential rotational velocity. The value of $(V/\sigma)_\mathrm{tot}$ remained at a similar value during the final \mbox{1.3 Myr} of the cluster assembly. With outward angular momentum transport, $(V/\sigma)_\mathrm{tot}$ is then expected to reduce later with time. The obtained $(V/\sigma)_\mathrm{tot}$ is in agreement with observations of GCs where typical values range between 0.1 and 0.6 in GCs that show statistically significant amounts of ordered rotation and that are dynamically young (relaxation time $>\mathrm{Gyr}$; e.g. \citealt{2018MNRAS.481.2125B}). Other definitions for $V/\sigma$, e.g. using the rotation amplitude over the total velocity dispersion observed in GCs in \citet{2025A&A...694A.184L}, also show similar values and similar correlation with dynamical age as in \citet{2018MNRAS.481.2125B}.

\begin{figure*}
\includegraphics[width=\textwidth]{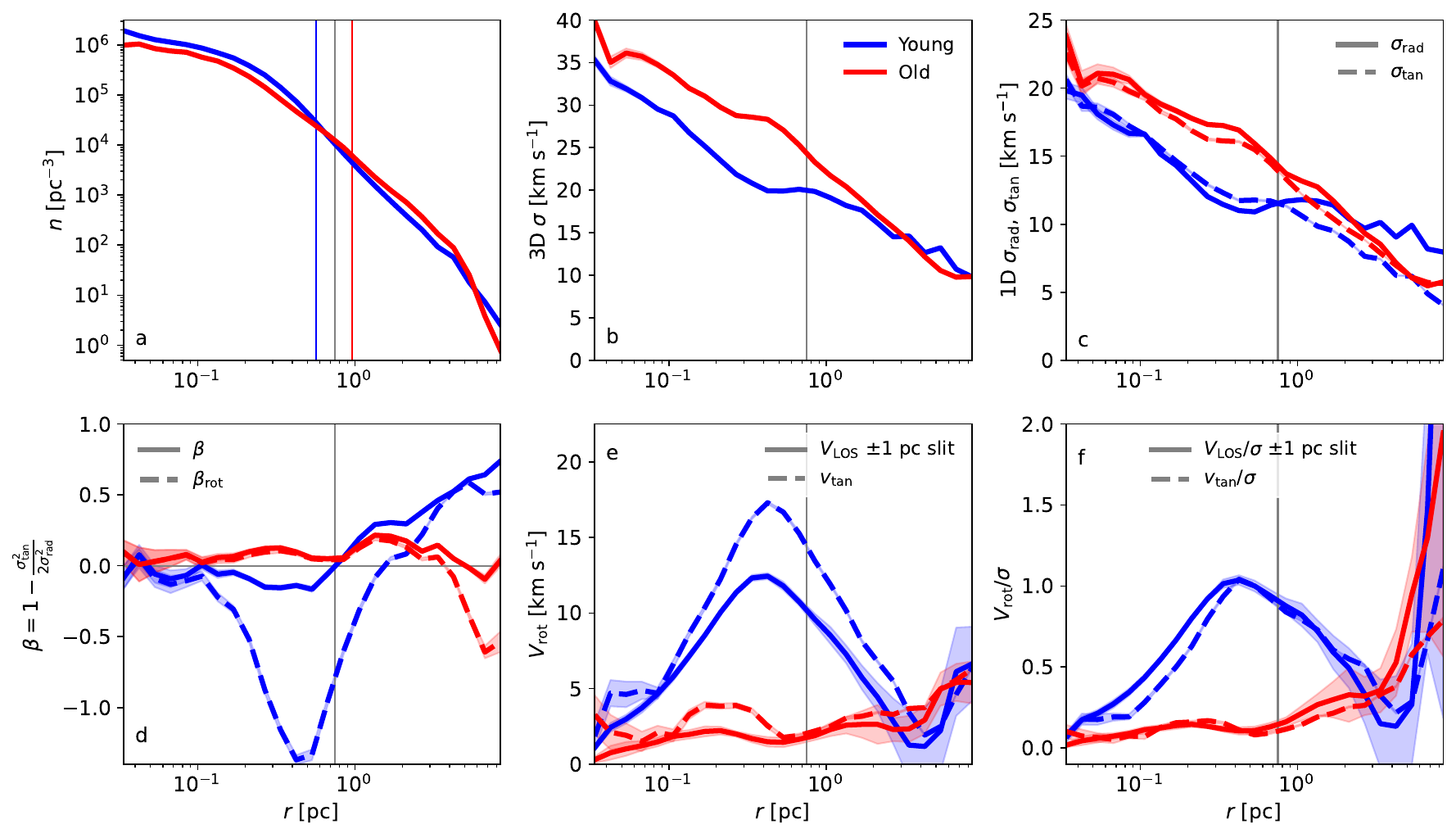}
\caption{The radial velocity distribution of the stars in the most massive cluster divided in to the younger (blue) and older (red) population in the final snapshot. The panels show the stellar number density (\textit{a}); 3D velocity dispersion (\textit{b}); radial and tangential velocity dispersion (\textit{c}); velocity anisotropy excluding ($\beta$, solid) and including ($\beta_\mathrm{rot}$, dashed) the bulk velocities (\textit{d}); the projected LOS rotation velocity in a $\pm 1$ pc slit and the tangential rotation velocity (solid and dashed lines, respectively; \textit{e}); and the corresponding $V/\sigma$ of the tangential and LOS velocity components (\textit{f}). The vertical gray line in the panels indicates the projected $R_\mathrm{eff}$, and the blue and red vertical lines in panel \textit{a} shows the values of $R_\mathrm{eff}$ for the young and old populations, respectively.
\label{fig:vels_pop}}
\end{figure*}

\begin{figure}
\includegraphics[width=\columnwidth]{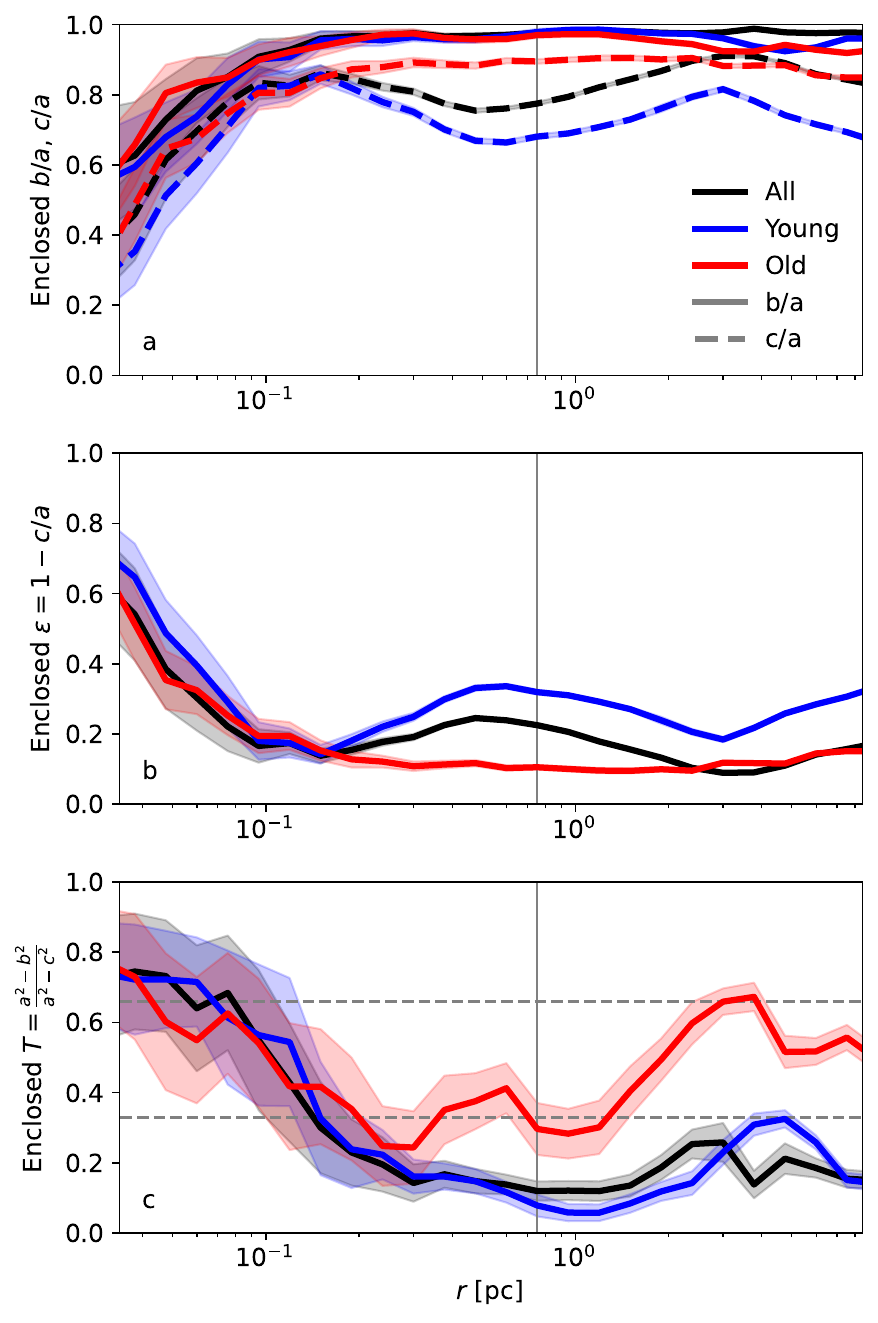}
\caption{The shape parameters of the most massive cluster in the final snapshot (black) divided additionally by age into two groups: stars younger (blue) and older (red) than the mean stellar age of the cluster. The panels show the axis ratios $b/a$ and $c/a$ (solid and dashed, respectively; panel \textit{a}), the ellipticity (panel \textit{b}) and the triaxiality (panel \textit{c}). All radial profiles are computed for stars enclosed within the increasing radial extent. The vertical lines show the projected $R_\mathrm{eff}$. The dashed horizontal lines in panel \textit{c} indicate the triaxiality values of $2/3$ and $1/3$ above and below which the shape is prolate and oblate, respectively.
\label{fig:shape}}
\end{figure}

\subsubsection{Kinematics of the age groups}\label{section:kinematics}

Finally we investigate whether the hierarchical cluster assembly leaves an imprint in the shape and kinematics of the stars. We divide the stars into two age groups either younger or older than the mean stellar age of the cluster in the final snapshot. The groups of stars have similar total masses within 7\%. The older population formed mainly in sub-clumps that accreted hierarchically before the cluster transitioned into the centrally concentrated star formation mode. The younger population corresponds to the latter.  To draw a qualitative parallel to the MPs of GCs, we may suppose that the formation of the chemically peculiar P2 occurs after the first VMSs have had time to enrich their surrounding, and thus the youngest stars in the cluster should also be the most chemically enriched. The age groups do not signify generations, but simply the two $\sim3$ Myr halves of the fairly Gaussian star formation history of this cluster that proceeded over a time span of $\sim6$ Myr (see Fig. 7 in \citealt{2024MNRAS.530..645L} for a similar star formation history). 

We compute the kinematic properties for these two populations in the same radial bins used for the whole stellar population. Fig. \ref{fig:vels_pop} shows the number density and velocity profiles of the two age groups. The young component is slightly more centrally concentrated with \mbox{$R_\mathrm{eff}\sim 0.56$ pc}, compared to \mbox{$R_\mathrm{eff}\sim 0.96$ pc} of the old component and \mbox{$R_\mathrm{eff}\sim 0.75$ pc} of the entire cluster. The young component has a significantly higher rotation velocity compared to the old component. The peak rotation rate of the old population of a few km s$^{-1}$ is consistent with the residual random angular momentum resulting from the hierarchical assembly process, based on initially non-rotating hierarchical N-body models of similar mass (\citealt{2024MNRAS.531.3770R, 2025arXiv250604330R}). The rotation velocity seen in Fig. \ref{fig:vels} (panels \textit{c} and \textit{d}) is therefore mainly dominated by the younger population of stars that has not yet relaxed unlike the old population. The radially biased anisotropy profile seen in Fig. \ref{fig:vels} (panel \textit{b}) at the outer parts is dominated by the younger component as well.

If we do not reduce the bulk velocity and only use the root mean square velocities to compute $\beta_\mathrm{rot}$, the young population shows at intermediate radii ($0.1$ pc $<r<2$ pc) a strong tangential bias (down to $\sim-1$) compared to the fairly isotropic profile at $r\lesssim R_\mathrm{eff}$ shown in Fig. \ref{fig:vels_pop}. This is simply a reflection of the rotation, that is encoded in bulk tangential motion. The radial bias in the outer radii of the young population is still retained even with bulk rotation included, as already shown in Fig. \ref{fig:vels}. The old population, on the other hand, remains isotropic when bulk rotation is not removed. The only exception is in the outermost parts where the $\beta_\mathrm{rot}$ of the old population shows tangential bias, i.e. bulk rotation.

$(V/\sigma)_\mathrm{tot}$ for the young and old populations are 0.6 and 0.1, respectively. Fig \ref{fig:vels_pop} (panel \textit{f}) shows the radial $V/\sigma$ profiles of the two components. The intermediate and inner radii show very different $V/\sigma$ shapes for the young and old populations in line with the differences in the rotation velocity profiles. In the outer radii, however, the populations show better agreement in their $V/\sigma$ profiles, reflecting their similar rotation rates at the outer radii. The only difference in the outer parts is the significant radial component in the velocity dispersion of the young population (as seen in $\beta$ and $\beta_\mathrm{rot}$), which is the result of young sub-clusters falling in on radial orbits.

In the absence of strong chemical variations as expected of MPs, another way of differentiating the sub-populations in addition to age would be based on their birth location. As the most massive stars are located in the core of the cluster, then this would be expected as the location where the most chemically enriched stars form. Our previous simulations indicate that the chemical enrichment does indeed occur concentrated in the central regions of massive clusters \citep{2024MNRAS.530..645L}. We tested performing the above analysis using a division into two populations in which one (here the second population) contains stars born within the half-mass radius of the most massive progenitor sub-cluster. The other population consists of stars that originate from the outer regions of the progenitor or from accreted sub-clumps (here the first population). The results for Fig. \ref{fig:vels_pop} remain qualitatively similar. The peak rotation rate of the inner population is reduced by 1--2 km s$^{-1}$ from that shown for the young population in panel $e$ of Fig. \ref{fig:vels_pop}. The rotation rate of the outer population peaks with $\sim5$ km s$^{-1}$ at $\sim 1$ pc. Some outer radial anisotropy is introduced into the outer population compared to the mostly isotropic old stars in panel $d$ of Fig. \ref{fig:vels_pop}. The most drastic change is, however, in the density profiles. The inner-formed population remains strongly centrally concentrated and dominates the density profile within $r_\mathrm{eff}$ even though the stars only constitute 1/3rd of the total cluster mass. To regain a less biased prediction of the density distributions, we chose to show only the populations divided by age in Fig. \ref{fig:vels_pop}. Stellar age can be viewed as a proxy for the progress of chemical enrichment in the cluster, and only selecting by age we allow later-formed stars also in the accreted sub-clusters to be considered a part of the population that would be analogous to P2 in observed GCs.

With this, we conclude that our results confirm the picture drawn out in previous simulation of star formation in pre-existing massive star clusters (\citealt{2010ApJ...724L..99B, 2021MNRAS.500.4578M, 2022MNRAS.517.1171L}) wherein the second formed population rotates significantly more than the first population. The previous simulations assumed a gradual initial rotation velocity for the first population, while here we self-consistently follow the formation and evolution of both two populations.

\subsubsection{Shapes of the age groups}\label{section:populations}

Finally we compute in 3D the axis ratios $b/a$ and $c/a$, ellipticity $\epsilon=1-c/a$ (i.e. maximal ellipticity given by the longest and shortest principal axes) and triaxiality parameter $T=\frac{a^2-b^2}{a^2-c^2}$ using the inertial tensor \citep{2008gady.book.....B} of the enclosed mass in increasing radial bins. The results are shown in Fig. \ref{fig:shape}.

Fig. \ref{fig:shape} shows the shape parameters for stars enclosed within increasing radii in the entire cluster and in the two age groups. The core of the cluster has a significant ellipticity (up to $\epsilon\sim0.6$) and a mildly prolate ($T>0.66$) or triaxial ($0.33<T<0.66$) shape reflecting the star formation out of the disky internal gas structure. The cluster becomes increasingly triaxial or oblate ($T<0.33$) when stars at outer radii are included, but note that the overall shape becomes more spherical e.g. with $\epsilon\sim 0.2$ within $R_\mathrm{eff}$. The older population drives down the ellipticity in the cluster as it is fairly triaxial/round, while the young population has a more elliptic shape throughout the cluster. 

Our new results verify the mildly elliptic shapes found for simulated young massive clusters in \citet{2020ApJ...904...71L}, where collisional dynamics was not yet considered. Observed (evolved) GCs show similarly low ellipticities \citep{Harris1996}, hence we conclude that massive dense clusters may have such spherical shapes already at birth.

\subsubsection{Comparison to NGC 104/47 Tuc}

The most massive  simulated young cluster has similar rotation characteristics with the Galactic GC NGC 104/47 Tuc that has been widely analysed in kinematic surveys (e.g. \citealt{2010MNRAS.406.2732L, 2017ApJ...844..167B, 2018MNRAS.481.2125B}). NGC 104 is already $\gtrsim11$ Gyr old \citep{2003A&A...408..529G, 2010ApJ...708..698D}, however it is still among the dynamically younger GCs \citep{2018MNRAS.478.1520B}. The differences in spatial and kinematic properties of our simulated young and old stellar populations (centrally more concentrated young stars; faster rotating young stars; radially biased outer young stars) are similar to what is found in some studies of NGC 104 that compare the chemically peculiar P2 and chemically normal P1 stars (P2 is more centrally concentrated; P2 rotates faster; P2 has a radial anisotropy in the outer parts; e.g. \citealt{2012ApJ...744...58M, 2024A&A...691A..94D, 2025A&A...694A.184L}; but see also e.g. \citealt{2023A&A...671A.106M} where no statistically significant difference between is found for P2 and P1). While the two age groups defined here are not directly analogous to the chemically identified multiple populations, the striking qualitative resemblance of the spatial and kinematic properties of the young and old population with the observed P2 and P1 in NGC 104 may signify a similar temporal sequence in their formation. In this scenario the P2 of NGC 104 would indeed have formed after P1 in a more centrally concentrated star formation phase. Previous cloud-scale simulations have also suggested that massive stars would form only late in the cluster assembly process \citep{2019MNRAS.490.3061V} supporting the late formation of the chemically peculiar population.

In some of the other dynamically young GCs, such as NGC 3201 and NGC 6101, the outer radii are also radially anisotropic, seemingly driven by the P2 stars \citep{2025A&A...694A.184L}. These clusters, however, show no clear differences in the rotation rates of the populations, and they have more centrally concentrated P1 populations as opposed to the more centrally concentrated P2 of NGC 104.

It remains to be tested whether the simulated populations retain their differences over an extended period of time. \citet{2013ApJ...779...85M} and \citet{2015MNRAS.450.1164H} for instance showed with N-body simulations that a disk-like young population can retain the input lower velocity dispersion and outer radial anisotropy for a long time. The specific internal dynamical evolution depends on the initial stellar distribution and kinematics, as well as on the evolution of the tidal field (e.g. \citealt{2025A&A...694A.163B}). The present results would thus offer more self-consistent initial conditions for follow-up N-body simulations.

\section{Conclusions and discussion}\label{section:conclusions}

We have presented the first galactic scale simulation of massive star cluster formation including collisional stellar dynamics in the vicinity of massive stars. The new model includes prescriptions for stellar collisions and TDEs that allow us to follow the collisional growth of VMSs in the dense cores of star clusters. 

The general population of very young ($\lesssim4$ Myr) intermediate mass star clusters (between a few $10^2$ \mdot{} and few $10^4$ \mdot) matches with the smallest sizes and highest surface densities of observed clusters younger than $10$ Myr in the LEGUS survey. The agreement is expected to further improve once the simulated clusters expand during their first few Myr of dynamical evolution (not included here). These clusters have on average twice the extent of similar clusters in a previous simulation in \citet{2024MNRAS.530..645L} that started with otherwise similar initial setup but did not implement collisional stellar dynamics. 

The most massive star cluster has a gravitationally bound mass of \mbox{$\sim2\times 10^5$ \mdot}. This is similar to what was found in the previous collisionless simulation. The \mbox{$\lesssim1$ pc} effective radius and surface density of $\sim 7\times 10^4$ \mdot{} pc$^{-2}$ in the most massive cluster approaches the extreme values of surface density and compactness ([$M_\mathrm{cluster}/10^5$ \mdot$]/[R_\mathrm{eff}/$ pc$]>1$) of observed young massive star clusters found in recent high-redshift data (e.g. \citealt{2024Natur.632..513A}). The density profiles increase toward the cluster centre, without a flat core. The central density (central $>1000$ stars) is more than two orders of magnitude higher and the central surface density an order of magnitude higher compared to the effective values. This result is consistent with the resolved best-fit profiles of \citet{2005ApJS..161..304M} where it is predominantly the dynamically young massive clusters (relaxation time $>$Gyrs) that show central-to-effective surface density ratios ($\Sigma_0/\Sigma_\mathrm{eff}$) of more than 10. The implication is that the GC progenitors observed at high-redshifts probably had at least an order of magnitude higher central densities than the resolution limited mean values of $\Sigma_\mathrm{eff}\sim10^5$--\mbox{$10^6$ \mdot{} pc$^{-2}$} reported thus far.

The most massive star cluster rotates at a rate of $V_\mathrm{LOS}\sim$\mbox{7 km s$^{-1}$} or $v_\mathrm{tan}\sim$\mbox{12 km s$^{-1}$}, with the radial rotation profile peaking inside the half-mass radius. To decipher which stars dominate the rotation, we further divided the stars of the cluster into a younger and an older half across the $\sim 6 $ Myr of star formation in the cluster. We find that the younger population is more centrally concentrated, it has a significantly faster peak rotation velocity (\mbox{$>12$ km s$^{-1}$} compared to 2--4 km s$^{-1}$ in the old population), and its velocity distribution is radially anisotropic in the outer parts. The first formed, older population is more isotropic and it has a larger velocity dispersion. The cluster has a significant elliptic and mildly prolate shape in the inner parts while the overall shape of the cluster is mildly elliptic ($\epsilon\sim 0.1$--$0.3$). The older population is more triaxial/round in shape throughout the cluster while the young population has a milddy prolate shape in the inner parts and an oblate shape in the outer parts. The first formed population has thus already relaxed throughout the hierarchical assembly process, while the younger population exhibits its initial shape and kinematics, including some radial motions in the outer parts from sub-clump infall. 

The radial stellar surface density and surface brightness profiles of the most massive simulated cluster are fairly similar in shape with the most massive local young star cluster R136. The shape and $V/\sigma$ values agree with dynamically young GCs. The kinematic properties of the young and old populations are in qualitatively agreement with those of the chemically peculiar P2 and normal P1 stars, respectively, in the Galactic GC NGC 104. In other dynamically young clusters, such as NGC 3201 and NGC 6101, the differences in the spatio-kinematics of the stellar populations can be less pronounced or even reversed compared to the features found here (e.g. a centrally more concentrated P1). Explaining the variety of population characteristics in GCs will thus require further investigation, including long-timescale N-body simulations.

Our stellar collision product of $\sim 1000$ \mdot{} would result in a black hole of at least $\sim600$ \mdot{} at our adopted metallicity of \mbox{$0.01$ \zdot}. The most massive star and the resulting black hole are therefore $\sim0.5\%$ of the total mass of the cluster. This is similar to the lower limits for the most massive stars found in N-body simulations where the collision products typically contain between a fraction of a per cent to $\sim 10\%$ of the cluster mass \citep{2004Natur.428..724P, 2023MNRAS.526..429A, 2024MNRAS.531.3770R, 2025arXiv250604330R}. Between a few per cent and $10\%$ is also the expected upper limit for extremely massive or supermassive stars forming through collisions or gas accretion based on analytic arguments \citep{2018MNRAS.478.2461G, 2025MNRAS.tmp.1257G}. We are therefore starting to probe the very high mass end of the stellar mass function, often found theoretically to extend to \mbox{$10^3$ \mdot{}} or \mbox{$10^4$ \mdot{}} (e.g. \citealt{2024Sci...384.1488F, 2025MNRAS.539.2561C}), now in galactic scale populations of simulated star clusters.

Our future aim is to explore the chemo-dynamical origin of multiple populations in GC progenitor clusters. While strong nitrogen enhancements are easily produced even with small amounts of hot hydrogen burning products, the currently implemented enrichment sources have not been able to produce large numbers of stars with as strong enhancements of, for instance, Al and Mg as observed (\citealt{2024MNRAS.530..645L}). This remains true even if we add the mass loss of up to $400$ \mdot{} from collisions and VMS winds related to the growth and evolution of the most massive star in the presently most massive cluster. Additional mass loss, for instance, through disruption of stars \citep{2017ApJ...836...80E} or the growth of the most massive star to the extreme or supermassive regime \citep{2014MNRAS.437L..21D, 2025MNRAS.tmp.1257G} will be needed to supply the star forming gas rapidly with significant amounts of light elements to explain the abundance anomalies in GCs. The growth could be enhanced through increased collision rates via primordial binaries \citep{2004MNRAS.352....1F, 2024ApJ...969...29G, 2025MNRAS.542L..78R} or with sustained gas accretion and associated mass loss \citep{2018MNRAS.478.2461G, 2025A&A...699A.223R}. The stellar evolution methodology can also be expanded in the future with more accurate criteria and non-mass conserving prescriptions for mass transfer and collisions (e.g. adopting the Roche lobe overflow criterion from \citealt{1983ApJ...268..368E}). The chemical enrichment due to TDEs \citep{2016MNRAS.458..127K} and binary evolution \citep{2009A&A...507L...1D, 2024ApJ...969...18N} may be explored with a method that returns the lost mass back into the ISM.

\section*{Acknowledgments}
The authors thank Angela Adamo, Ellen Leitinger, Michael Hilker and Holger Baumgardt for clarifying discussions about the observational data. NL and TN gratefully acknowledge the Gauss Centre for Supercomputing e.V. (www.gauss-centre.eu) for funding this project by providing computing time on the GCS Supercomputer SUPERMUC-NG at Leibniz Supercomputing Centre (www.lrz.de) under project numbers pn49qi. TN acknowledges support from the Deutsche Forschungsgemeinschaft (DFG, German Research Foundation) under Germany's Excellence Strategy - EXC-2094 - 390783311 from the DFG Cluster of Excellence "ORIGINS". The computations were carried out at SuperMUC-NG hosted by the LRZ and the FREYA cluster hosted by The Max Planck Computing and Data Facility (MPCDF) in Garching, Germany.

This research made use of \textsc{python} packages \textsc{scipy} \citep{2020SciPy-NMeth}, \textsc{numpy} \citep{2020NumPy-Array}, \textsc{matplotlib} \citep{Hunter:2007}, \textsc{pygad} \citep{2020MNRAS.496..152R}, \textsc{h5py} \citep{collette_python_hdf5_2014}, and \textsc{astropy}\footnote{http://www.astropy.org} \citep{2022ApJ...935..167A}.

\section*{Data availability statement}
The data relevant to this article will be shared on reasonable request to the corresponding author.


\bibliographystyle{mnras}
\interlinepenalty=10000



\bsp	
\label{lastpage}
\end{document}